\documentclass[twocolumn,preprintnumbers]{revtex4}
\usepackage{amssymb}
\usepackage{amsmath}
\usepackage{amsfonts}
\usepackage{graphicx}
\usepackage{dcolumn}
\usepackage{color}
\usepackage{bm}

\begin{document}

\title{Ultrafast molecular dynamics in terahertz-STM experiments: \\
Theoretical analysis using Anderson-Holstein model }
\author{Tao Shi$^{1}$, J. Ignacio Cirac$^{2,3}$, and Eugene Demler$^{4}$}
\affiliation{$^{1}$ Institute of Theoretical Physics, Chinese Academy of Sciences, P.O.
Box 2735, Beijing 100190, China \\
$^{2}$Max-Planck-Institut f\"{u}r Quantenoptik, Hans-Kopfermann-Strasse. 1,
85748 Garching, Germany \\
$^{3}$Munich Center for Quantum Science and Technology (MCQST),
Schellingstr. 4, D-80799 M\"{u}nchen, Germany \\
$^{4}$Department of Physics, Harvard University, 17 Oxford st., Cambridge,
MA 02138}
\date{\today }

\begin{abstract}
We analyze ultrafast tunneling experiments in which electron transport
through a localized orbital is induced by a single cycle THz pulse. We
include both electron-electron and electron-phonon interactions on the
localized orbital using the Anderson-Holstein model and consider two
possible filling factors, the singly occupied Kondo regime and the doubly
occupied regime relevant to recent experiments with a pentacene molecule.
Our analysis is based on variational non-Gaussian states and provides the
accurate description of the degrees of freedom at very different energies,
from the high microscopic energy scales to the Kondo temperature $T_K$. To
establish the validity of the new method we apply this formalism to study
the Anderson model in the Kondo regime in the absence of coupling to
phonons. We demonstrate that it correctly reproduces key properties of the
model, including the screening of the impurity spin, formation of the
resonance at the Fermi energy, and a linear conductance of $2e^2/h$. We
discuss the suppression of the Kondo resonance by the electron-phonon
interaction on the impurity site. When analyzing THz STM experiments we
compute the time dependence of the key physical quantities, including
current, the number of electrons on the localized orbital, and the number of
excited phonons. We find long-lived oscillations of the phonon that persist
long after the end of the pulse. We compare the results for the interacting
system to the non-interacting resonant level model.
\end{abstract}

\maketitle

\section{Introduction and Model}

\subsection{Motivation}

\begin{figure}[htp]
\begin{center}
\includegraphics[width=0.4\textwidth]{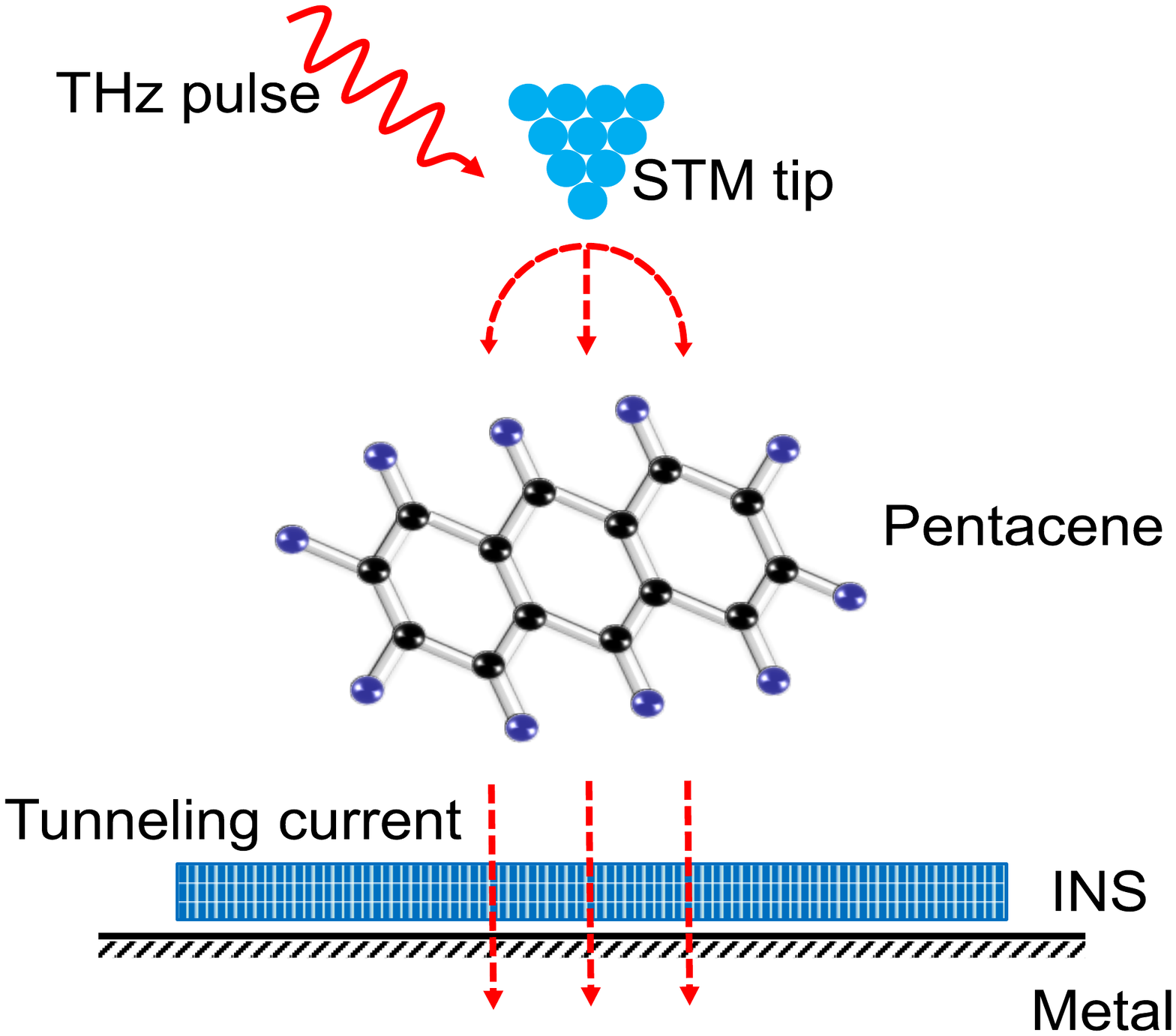}
\end{center}
\caption{Light-assistend tunneling through a single molecule: schematic of a
THz-STM experiment.}
\label{setup_fig}
\end{figure}

Ultrafast experiments constitute a new approach to exploring quantum
many-body systems and provide a platform for developing new types of solid
state devices for nanotechnology and quantum information processing (for
review see Refs. \cite{Basov2011,Kampfrath2013,Giannetti2016,Basov2017}).
One of the promising techniques is a recently developed terahertz STM
(THz-STM) that integrates femtosecond lasers with scanning tunneling
microscopes (STM) \cite{Cocker2013,Yoshioka2016,Cocker2016,Jelic2017}. This
technique allows to combine atomic spatial resolution of STM with
subpicosecond coherent temporal control of electron currents. Such
experiments pose a new challenge to many-body theory to develop methods for
analyzing the far out of equilibrium quantum dynamics of interacting
many-body systems. Motivated by recent experiments in this paper we provide
a theoretical analysis of THz-STM experiments of tunneling through a single
localized orbital, such as a HOMO orbital in a pentacene molecule used by
Cocker \textit{et al}. \cite{Cocker2016} (see Fig. \ref{setup_fig}). Our
analysis extends earlier theoretical studies of such systems (see \cite%
{Cuevas1990,Galperin2007} and references therein) by including a
non-perturbative treatment of electron-phonon and electron-electron
interactions.

\begin{figure}[htp]
\begin{center}
\includegraphics[width=0.4\textwidth]{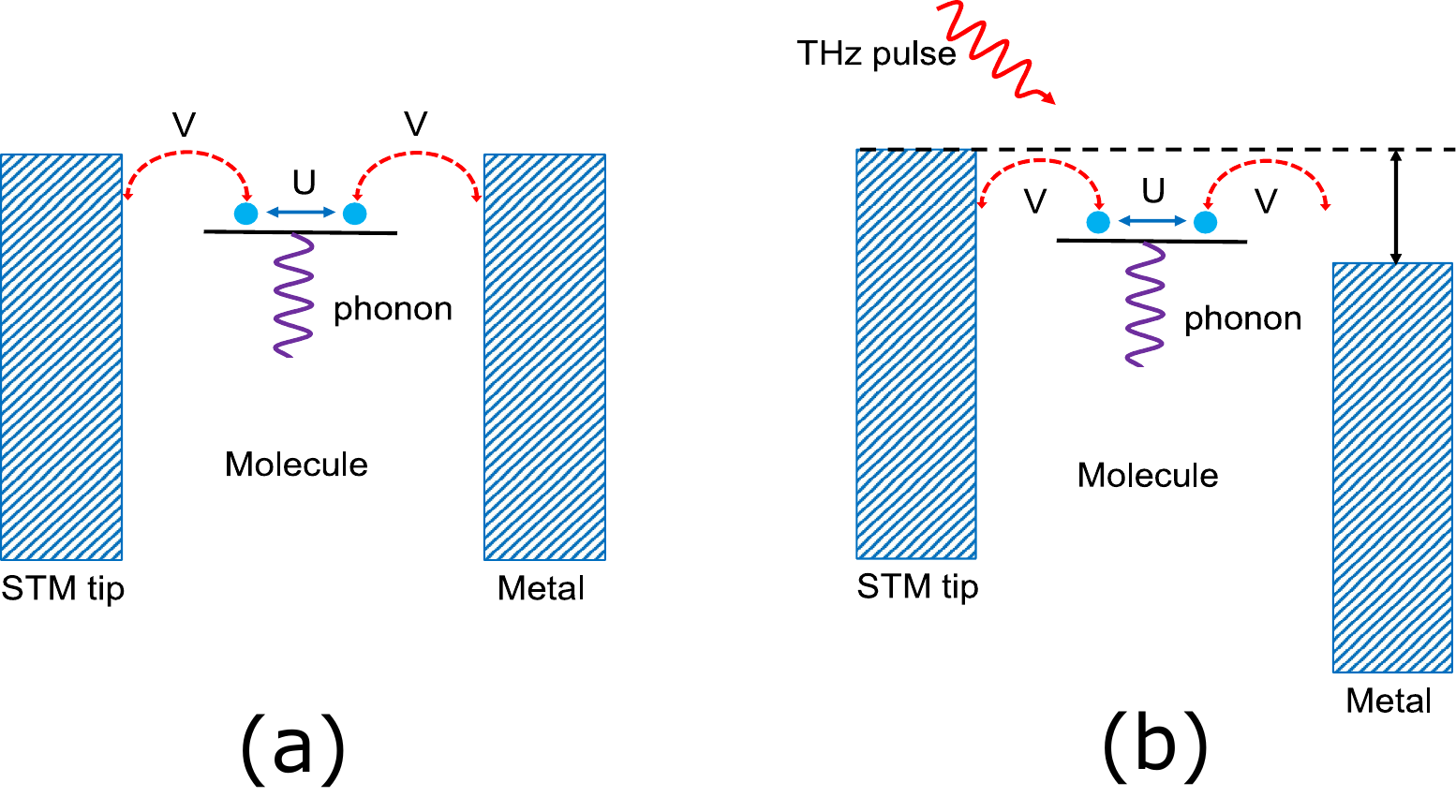}
\end{center}
\caption{(a) Anderson-Holstein microscopic model for tunneling through a
molecule. Only the HOMO orbital on the molecule is considered. Coulomb
repulsion between electrons on the molecular orbital is included through the
Anderson model. Interaction of electrons with the optical vibrational mode
is included using the Holstein model. (b) Light assisted tunneling through a
molecule in THz-STM experiments. Electric field from the femtosecond pulse
modifies energy differences between the reservoirs and the molecular level.}
\label{Anderson}
\end{figure}

In writing this paper we set ourselves two goals. Firstly, we present a new
theoretical approach for analyzing non-equilibrium dynamics of the
Anderson-Holstein impurity model. This method is based on the variational
non-Gaussian wavefunctions introduced in Ref. \cite{Shi2018} and further
extended in the context of quantum impurity models in Refs. \cite%
{Ashida2018,Ashida2018a}. This technique is versatile and can be applied to
a broad class of nonequilibrium problems, including quenches, pump and probe
experiments, and the analysis of AC and DC transport. Most of the previously
developed approaches for the Anderson impurity problem focus either on high
energy degrees of freedom on the scale of the electron-electron repulsion $U$
\cite{Ng1988}, or the low energy sector with the scale set by the Kondo
temperature $T_{K}$ \cite{Wingreen1993,Wingreen1994,Ng1996,Kaminski2000}.
The advantage of our approach is that within the same framework it describes
both high and low energy degrees of freedom without requiring numerical
resources of NRG \cite{Hewson2002,Legeza2008,Holzner2010} or DMRG \cite%
{Al-Hassanieh2006,Heidrich-Meisner2009,Eckel2010}. To establish the validity
of the new method we verify that it correctly reproduces basic results of
the canonical Anderson impurity model: it captures the Kondo resonance in
the spectral function at the Fermi energy and gives a conductance $%
G=2e^{2}/h $ in the linear response regime. We extend the analysis of the
Anderson model to include the electron-phonon interaction on the impurity
site and demonstrate that coupling to the phonons can strongly suppress the
Kondo peak. Our second goal in this paper is to analyze a specific type of
nonequilibrium dynamics: THz STM experiments through a single molecule. For
such experiments we compute the time dependence of the key observables:
current through the system, number of electrons on the molecule, and number
of excited phonons. To illustrate the important role of interactions in the
system, we also discuss a non-interacting resonant level model (RLM)in the
infinite bandwidth approximation. We show that, in this case, the transient
current dynamics in the THz-STM experiemnts can be computed analytically. We
discuss the difference in the results between the non-interacting RLM system
and the Anderson-Holstein model.

\subsection{Anderson-Holstein Model}

The system we consider is shown schematically in Figs. \ref{setup_fig} and %
\ref{Anderson}. In the absence of laser light the chemical potential of the
organic molecule is in the middle of the gap between the HOMO and LUMO
orbitals of the molecule and there is no current through the system \cite%
{Cocker2016}. The light pulse changes the relative energies of the molecular
level and the reservoirs (see Fig. \ref{Anderson}b) and allows for an
ultrafast electron burst as a co-tunneling process between the STM tip, the
molecule, and the metal. Due to the asymmetry of the electric field in the
pulse \cite{Cocker2013,Yoshioka2016} it is sufficient to consider only one
of the orbitals in the molecule, which we take to be the HOMO orbital. We
model the interaction between HOMO electrons using the Anderson type local
repulsion $U$ and include interaction of electrons with vibrations of the
molecule \cite{Cocker2016} using Holstein type coupling of the phonon
displacement to the number of electrons in the localized orbital \cite%
{Mahan2000}. Thus, molecular degrees of freedom are given by the Hamiltonian
\begin{eqnarray}
\mathcal{H}_{\mathrm{mol}} &=&\mathcal{H}_{\mathrm{A}}+\mathcal{H}_{\mathrm{%
phon}},  \notag \\
\mathcal{H}_{\mathrm{A}} &=&\varepsilon _{d}n_{d}+Un_{\uparrow
}n_{\downarrow },  \notag \\
\mathcal{H}_{\mathrm{phon}} &=&\omega _{b}b^{\dagger
}b+g(2-n_{d})(b^{\dagger }+b).
\end{eqnarray}%
Here, $d_{\sigma }$ are the annihilation operators of electrons in the HOMO,
$n_{d}=n_{\uparrow }+n_{\downarrow }$ is the total number of electrons in
the molecule $n_{\sigma =\uparrow \downarrow }=d_{\sigma }^{\dagger
}d_{\sigma }$, $b$ is the annihilation operator for the phonon with
frequency $\omega _{b}$, and $\varepsilon _{d}$ is the energy of the HOMO
orbital.

We model the STM tip and the metallic electrode as one dimensional chains of
non-interacting electrons because we do not expect appreciable dependence on
the nature of electron reservoirs. Our analysis can be easily extended to
other type of reservoirs.
\begin{equation}
\mathcal{H}_{\mathrm{a}=\mathrm{L,R}}=-t_{0}\sum_{\left\langle
ij\right\rangle ,\sigma }c_{i,\sigma ,\mathrm{a}}^{\dagger }c_{j,\sigma ,%
\mathrm{a}}-\mu _{\mathrm{a}}\sum_{j,\sigma }c_{j,\sigma ,\mathrm{a}%
}^{\dagger }c_{j,\sigma ,\mathrm{a}},
\end{equation}%
Here, $c_{j,\sigma ,\mathrm{a}}$ denote annihilation operators of electrons
with spin $\sigma $ in the reservoir $\mathrm{a}=\mathrm{L,R}$ at site $j$.
Coupling between the molecule and reservoirs is included via the
hybridization term%
\begin{equation}
\mathcal{H}_{\mathrm{V}}=V\sum_{\sigma ,\mathrm{a}}(c_{0,\sigma ,\mathrm{a}%
}^{\dagger }d_{\sigma }+\mathrm{H.c.}).
\end{equation}%
Different chemical potentials of the reservoirs account for a finite
time-dependent bias voltage, where we assume that all the time dependent
potential is applied between the molecule and the right reservoir (our
analysis can be easily generalized to the arbitrary time dependent
potentials for both reservoirs)
\begin{equation}
V_{\mathrm{e}}(t)=\mu _{\mathrm{R}}(t)-\mu _{\mathrm{L}}.
\end{equation}%
The main effect of the femtosecond laser pulse is to change the bias voltage
as shown in Fig. \ref{Anderson}b.

The full Hamiltonian of the system is then given by
\begin{eqnarray}
\mathcal{H} (t) = \mathcal{H}_{\mathrm{mol}} + H_{\mathrm{V}} + \sum_a
\mathcal{H}_{\mathrm{a}} (t)
\end{eqnarray}

\subsection{Review of the theoretical formalism}

The theoretical modeling of light assisted tunneling of electrons through a
single molecule requires the analysis of nonequilibrium dynamics of the
Anderson impurity model with an added complication, that of the
electron-phonon coupling. The Anderson impurity problem is one of the most
fundamental models in the field of interacting electron systems. It provides
the foundation for our understanding of a broad range of physical phenomena,
including the Kondo effect in metals \cite{Mahan2000,Hewson1993}, the
electron transport in mesoscopic structures \cite%
{Kastner1998,Cronenwett1998,Kouwenhoven2001,Bolech2016}, the formation of
heavy fermion electron systems \cite%
{Stewart1979,Gegenwart2008,Si2010,Kotliar2004}. A broad range of analytical
tools has been developed to study equilibrium properties of this model
including a Bethe ansatz solution \cite{Andrei1983,Wiegmann1984}, the
renormalization group approach \cite{Bulla2008}, the slave-particle method
\cite{Read1983,Kroha1999}, and dynamical mean-field theory (DMFT) \cite%
{Georges1996,Feldbacher2004,Rubtsov2005,Werner2006,Gull2011}. However, the
non-equilibrium dynamics of the model remains poorly understood. While most
of the theoretical work on quantum dynamics of the Anderson model relied on
the non-crossing approximation \cite{Wingreen1994,Anders1994}, other
promising new approaches utilized real-time DMFT \cite{Aoki2014} combining
the bold-line quantum Monte Carlo technique with a memory function formalism
\cite{Cohen2013}. Calculations using either of these approaches are
demanding in terms of numerical resources.

In this paper we propose a new approach to analyzing the Anderson-Holstein
impurity model both in and out of equilibrium. This approach is based on
combining a unitary transformation of the generalized-polaron type with a
family of non-Gaussian variational states for quantum impurity problems. The
electron-phonon part of the transformation entangles the electron and phonon
degrees of freedom and allows us to use factorized wavefunctions, in which
bosonic and electronic parts are described using a Gaussian ansatz and a
family of non-Gaussian variational wavefunctions, respectively. The role of
the unitary transformation in the non-Gaussian ansatz for the Anderson model
is to utilize exact conservation laws to reduce the number of impurity
degrees of freedom at the cost of introducing additional interactions and
correlations between the bath particles. The key difference between our
method and the traditional approaches based on the polaron transformations
\cite{Mahan2000} is that we allow parameters of the transformation to vary
in time. The general philosophy of this method has been introduced earlier
in Ref. \cite{Shi2018}, which demonstrated that this technique successfully
describes equilibrium and dynamical properties of many important solid state
systems including polarons in the Su-Schrieffer-Heegger and Holstein models,
spin-bath models, and superconductivity in the Holstein model. Recently,
Ashida \textit{et al}. \cite{Ashida2018,Ashida2018a} demonstrated that this
method efficiently captures the complicated dynamics of the anisotropic
Kondo model, including the finite time crossover between the ferromagnetic
and antoferromagnetic couplings \cite{Kanasz-Nagy2018}. Since this is the
first time that this method is used to study the Anderson model, we include
a separate analysis of the equilibrium properties including the electron
spectral function. We demonstrate that our approach accurately captures the
formation of the Kondo resonance at the Fermi energy with the width set by
the Kondo energy scale. This demonstrates that our method is versatile and
captures both short time/high energy and long time/low energy aspects of the
Anderson model.

\subsection{Organization of the paper}

This paper is organized as follows: In Section II we introduce the general
formalism of non-Gaussian variational wavefunctions. In Section III we use
the imaginary time flow approach to analyze the ground state properties of
the Anderson and Anderson-Holstein models focusing on the regimes of single
and double occupancy of the localized orbital. We compute electron spectral
functions and demonstrate that in the Kondo regime of the Anderson model our
approach correctly reproduces a resonance at the Fermi level. We show how
the spectral functions get modified upon adding the interaction with
phonons. Two prominent features are the suppression of the Kondo resonance
and the appearance of the phonon shake-off peaks. As additional check of the
validity of our method we compute spin correlation functions between
electrons on the localized orbital and in the electron reservoirs. We
demonstrate that in the Kondo regime we get the expected oscillating
correlations decaying as a power law of the distance to the localized
orbital. In Section IV we introduce a real time analysis of the Anderson and
Anderson-Holstein models. For the Anderson model we analyze the DC transport
and compute the full non-linear conductance. We demonstrate that in the
Kondo regime our method correctly gives a linear conductance equal to $%
2e^{2}/h$. For the Anderson-Holstein model, we focus on the analysis of
photocurrent induced by the THz pulze in the single and double occupancy
regimes. We present results for the time evolution of the current, the
occupation number of the localized orbital, and the phonon displacement. We
show the dependence of the total transferred charge on the amplitude of the
electromagnetic pulse. Section V contains a summary of our results and a
discussion of interesting open problems. Many technical details of the
calculations are not presented in the main text of the paper but delegated
to Appendices. This includes a derivation of the analytical results for
photocurrent in the non-interacting resonant level model.

\section{Non-Gaussian variational ansatz}

In this section, we introduce the variational ansatz%
\begin{equation}
\left\vert \Psi _{\mathrm{NGS}}\right\rangle =U_{\mathrm{ph}}U_{\mathrm{A}%
}\left\vert \Psi _{\mathrm{GS}}\right\rangle _{f}\left\vert \Psi _{\mathrm{GS%
}}\right\rangle _{b},  \label{NGS}
\end{equation}%
which we will use to study both the ground state and non-equilibrium
real-time dynamics. The unitary transformation $U_{\mathrm{ph}}$ entangles
the phonon mode with the electronic degrees of freedom in HOMO and in the
reservoirs. The unitary transformation $U_{\mathrm{A}}$ uses parity
conservation to partially decouple the impurity degrees of freedom.
Throughout the paper we use the terminology of the Anderson impurity model
and refer to the electronic degrees of freedom in the molecule as impurity
degrees of freedom. Finally $\left\vert \Psi _{\mathrm{GS}}\right\rangle
_{f,b}$ are Gaussian states for electrons and the phonon mode.

\subsection{Electron-phonon polaron transformation}

The generalized polaron transformation $U_{\mathrm{ph}}=e^{S}$ is defined by
the generating operator
\begin{equation}
S=iR^{T}\lambda (2-n_{d})  \label{S_polaron}
\end{equation}%
with the time-dependent variational parameters $\lambda =(\lambda
_{x},\lambda _{p})^{T}$. Here $R=(x,p)=(b^{\dagger }+b,i(b^{\dagger
}-b))^{T} $ is the quadrature of the phonon mode. When the polaron
transformation (\ref{S_polaron}) is applied to the system Hamiltonians
\begin{equation}
H_{1}=U_{\mathrm{ph}}^{\dagger }HU_{\mathrm{ph}}
\end{equation}%
we find that the lead Hamiltonians $\mathcal{H}_{\mathrm{L,R}}$ do not
change, whereas the molecular part of the Hamiltonian $\mathcal{H}_{\mathrm{%
mol}}=\mathcal{H}_{\mathrm{A}}+\mathcal{H}_{\mathrm{phon}}$ becomes%
\begin{eqnarray}
\mathcal{H}_{\mathrm{A}} &=&\tilde{\varepsilon}_{d}n_{d}+\tilde{U}%
n_{\uparrow }n_{\downarrow }+4(\omega _{b}\lambda ^{T}\lambda -2g\lambda
_{p}),  \notag \\
\mathcal{H}_{\mathrm{phon}} &=&\frac{1}{4}\omega
_{b}R^{T}R+(2-n_{d})R^{T}G_{\lambda },
\end{eqnarray}%
and the hybridization term changes to%
\begin{equation}
H_{\mathrm{V}}=Ve^{-iR^{T}\lambda }\sum_{\sigma ,\mathrm{a}}c_{0,\sigma ,%
\mathrm{a}}^{\dagger }d_{\sigma }+\mathrm{H.c.}.
\end{equation}

The polaron transformation reduces the electron-phonon interaction to $%
G_{\lambda }=\left( g-\omega _{b}\lambda _{p},\omega _{b}\lambda _{x}\right)
^{T}$, renormalizes the single particle energy level $\tilde{\varepsilon}%
_{d}=\varepsilon _{d}-3(\omega _{b}\lambda ^{T}\lambda -2g\lambda _{p})$ and
the on-site interaction $\tilde{U}=U+2(\omega _{b}\lambda ^{T}\lambda
-2g\lambda _{p})$, whereas the hybridization term $Ve^{-iR^{T}\lambda }$
between the phonon-dressed electrons in HOMO and reservoirs aquires
polaronic dressing.

\subsection{Parity operator and impurity decoupling transformation}

Refs \cite{Shi2018,Ashida2018} introduced an efficient way of solving
quantum impurity problems by utilizing parity conservation. In particular,
in the Kondo impurity problem one can construct an exact unitary
transformation that maps the conserved parity operator to one of the
components of the impurity spin. After the transformation, the impurity spin
is decoupled from the reservoir at the cost of introducing interactions
between bath fermions, which, however, can be effectively handled using the
Gaussian part of the wavefunction. In this paper we generalize such
transformation to the Anderson impurity problem. In the Anderson model one
can not completely decouple the impurity from the reservoir, but it is
possible to reduce the number of degrees of freedom so that the Gaussian
ansatz for the wavefunction can be utilized efficiently.

To simplify notations we introduce a four component representation of the
electronic Hilbert space on the molecule: $\{\left\vert
0\right\rangle,\left\vert \uparrow \right\rangle = d_\uparrow^\dagger | 0
\rangle ,\left\vert \downarrow \right\rangle = d_\downarrow^\dagger | 0
\rangle,\left\vert \uparrow \downarrow \right\rangle = d_\uparrow^\dagger
d_\downarrow^\dagger | 0 \rangle\}$. We define four component matrices in
this space so that the first Pauli matrix acts in the space (1,2) $%
\leftrightarrow$ (3,4). For example,
\begin{eqnarray}
\sigma_x \otimes I = \left[
\begin{array}{cc}
\hat{0} & \hat {1} \\
\hat{1} & \hat{0}%
\end{array}
\right] \hspace{1cm} \sigma_z \otimes I = \left[
\begin{array}{cc}
\hat{1} & \hat {0} \\
\hat{0} & \hat{-1}%
\end{array}
\right]
\end{eqnarray}
When the Pauli matrix is in the second position in the tensor product, it
acts simultaneously in sub-spaces 1 $\leftrightarrow$ 2, 3 $\leftrightarrow$
4, so that
\begin{eqnarray}
I \otimes \sigma_x = \left[
\begin{array}{cc}
\hat{\sigma_x} & \hat {0} \\
\hat{0} & \hat{\sigma_x}%
\end{array}
\right] \hspace{1cm} I \otimes \sigma_z = \left[
\begin{array}{cc}
\hat{\sigma_z} & \hat {0} \\
\hat{0} & \hat{\sigma_z}%
\end{array}
\right]
\end{eqnarray}

We define the parity operator for spin $\uparrow $ electrons on the molecule
\begin{equation*}
\Sigma _{z}=e^{i\pi (d_{\uparrow }^{\dagger }d_{\uparrow }-1)}
\end{equation*}%
The choice of the overall sign is a matter of convenience. In the tensor
notations introduced earlier
\begin{equation*}
\Sigma ^{z}=-I\otimes \sigma ^{z}
\end{equation*}%
Note that the operator $\Sigma _{z}$ performs a $\pi $ rotation in the
subspace ($|\uparrow \rangle $, $|\downarrow \rangle $) as well as a $\pi $
rotation in the subspace ($|0\rangle $, $|\uparrow \downarrow \rangle $).
The parity operator for spin $\uparrow $ fermions in the bath is defined as
\begin{eqnarray}
P_{z} &=&e^{i\pi N_{\uparrow }} \\
N_{\uparrow } &=&\sum_{j,\mathrm{a}}c_{j,\uparrow ,\mathrm{a}}^{\dagger
}c_{j,\uparrow ,\mathrm{a}}
\end{eqnarray}%
The parity operator for the system as a whole
\begin{equation*}
P=\Sigma ^{z}P_{z}
\end{equation*}%
is conserved since the Hamiltonian preserves the number of electrons with a
given spin. Mathematically this means that $[P,H_{1}]=0$.

In the spirit of Refs\cite{Ashida2018,Ashida2018a} we introduce a
transformation that maps the conserved operator $P$ entirely into the
impurity degrees of freedom. Following this transformation, the original
conservation low for $P$ becomes a conservation law for the imourity
electrons. We consider a unitary operator
\begin{equation}
U_{\mathrm{A}}=\frac{1}{\sqrt{2}}(1+i\Sigma ^{y}P_{z})
\label{U_A_definition}
\end{equation}%
with $\Sigma ^{y}=-\sigma ^{x}\otimes \sigma ^{y}$ Direct calculation shows
that
\begin{equation}
U_{\mathrm{A}}^{\dagger }PU_{\mathrm{A}}=\sigma ^{x}\otimes \sigma
^{x}\equiv X
\end{equation}%
We observe that the operator $X$ does not contain any degrees of freedom of
the reservoir electrons. Another important feature of the operator $X$ is
that it only connects states with the same electron parity, i.e. it does not
mix between subspaces ($|\uparrow \rangle $, $|\downarrow \rangle $) and ($%
|0\rangle $, $|\uparrow \downarrow \rangle $). Hence the operator $X$ is
bosonic and its eigenstates are physically well defined states. After
performing the transformation $U_{\mathrm{A}}$ on the Hamiltonian
\begin{equation*}
H_{2}=U_{\mathrm{A}}^{\dagger }H_{1}U_{\mathrm{A}}
\end{equation*}%
we are guaranteed that the operator $X$ commutes with the transformed
Hamiltonian $H_{2}$, which means that we reduced the number of degrees of
freedom corresponding to the molecule. Let us discuss the implications of
this fact in more detail. Operator $X$ has eigenvalues $\pm 1$. The
eigenstates corresponding to the eigenvalue $+1$ make a two dimensional
Hilbert space with basis vectors
\begin{eqnarray}
|+_{s}\rangle &=&\frac{1}{\sqrt{2}}(|\uparrow \rangle +|\downarrow \rangle )
\notag \\
|+_{c}\rangle &=&\frac{1}{\sqrt{2}}(|0\rangle +|\uparrow \downarrow \rangle )
\label{plus_states}
\end{eqnarray}%
The eigenstates corresponding to the eigenvalue $-1$ make an orthogonal two
dimensional Hilbert space with basis vectors
\begin{eqnarray}
|-_{s}\rangle &=&\frac{1}{\sqrt{2}}(|\uparrow \rangle -|\downarrow \rangle )
\notag \\
|-_{c}\rangle &=&\frac{1}{\sqrt{2}}(|0\rangle -|\uparrow \downarrow \rangle )
\label{minus_states}
\end{eqnarray}%
Conservation of $X$ means that the dynamics described by the Hamiltonian $%
H_{2}$ can not mix between subspaces (\ref{plus_states}) and (\ref%
{minus_states}). Hence for the dynamics in the $+1$ subspace we need to
consider only two electronic states in the molecule: $|+_{s}\rangle $ and $%
|+_{c}\rangle $. We introduce the fermion operator
\begin{equation*}
f_{+}=(-)^{F}\,|+_{s}\rangle \langle +_{c}|
\end{equation*}%
where $F=\sum_{j,\sigma ,\mathrm{a}}c_{j,\sigma ,\mathrm{a}}^{\dagger
}c_{j,\sigma ,\mathrm{a}}$ is the total number operator of electrons in both
reservoirs. In the $+1$ subspace Hamiltonian $H_{2}$ can be expressed
entirely in terms of the reservoir operators $c_{i,\sigma ,a}$, for the
molecule $f_{+}^{\pm }$, and phonon operators. Analogously, in the $-1$
subspace (\ref{minus_states}) we introduce the fermion operator that
connects two states with different electron parity
\begin{equation*}
f_{-}=(-)^{F}\,|-_{s}\rangle \langle -_{c}|
\end{equation*}

The explicit expression for $H_{2}$ can be obtained from a straightforward
but somewhat lengthy calculation. The part of the Hamiltonian corresponding
to the left and right reservoirs does not change. The part of $H_{2}$
corresponding to the molecule, $\mathcal{H}_{\mathrm{mol}}=\mathcal{H}_{%
\mathrm{A}}+\mathcal{H}_{\mathrm{phon}}$, becomes%
\begin{eqnarray}
H_{\mathrm{A}} &=&4(\omega _{b}\lambda ^{T}\lambda -2g\lambda _{p})+\tilde{%
\varepsilon}_{d}  \notag \\
&&+\frac{1}{2}\tilde{U}f_{\gamma }^{\dagger }f_{\gamma }+(\tilde{\varepsilon}%
_{d}+\frac{1}{2}\tilde{U})\gamma P_{z}f_{\gamma }^{\dagger }f_{\gamma },
\notag \\
H_{\mathrm{ph}} &=&\frac{1}{4}\omega _{b}R^{T}R+(1-\gamma P_{z}f_{\gamma
}^{\dagger }f_{\gamma })R^{T}G_{\lambda },
\end{eqnarray}%
The transformed hybridization term is given by%
\begin{eqnarray}
H_{\mathrm{V}} &=&\frac{V}{2}e^{-iR^{T}\lambda }\sum_{\mathrm{a}%
}[c_{0,\uparrow ,\mathrm{a}}^{\dagger }(f_{\gamma }^{\dagger }+f_{\gamma
})+\gamma c_{0,\downarrow ,\mathrm{a}}^{\dagger }(f_{\gamma }^{\dagger
}-f_{\gamma })  \notag \\
&&-P_{z}(\gamma c_{0,\uparrow ,\mathrm{a}}^{\dagger }+c_{0,\downarrow ,%
\mathrm{a}}^{\dagger })(f_{\gamma }^{\dagger }+f_{\gamma })]+\mathrm{H.c.},
\end{eqnarray}

\subsection{Gaussian part of the wavefunction}

A convenient way of defining Gaussian wavefunctions in equation (\ref{NGS})
is to consider a unitary Gaussian transformation acting on the electron and
phonon vacuum $\left\vert \Psi _{\mathrm{GS}}\right\rangle _{f}\left\vert
\Psi _{\mathrm{GS}}\right\rangle _{b}=U_{\mathrm{GS}}\left\vert
0\right\rangle $. The Gaussian state is completely characterized by the
expectation values of the phonon quadratures $\Delta _{R}=\left\langle
R\right\rangle $ and covariance matrices $\Gamma _{b}=\left\langle \{\delta
R,\delta R^{T}\}\right\rangle _{\mathrm{GS}}/2$ and $\Gamma
_{f}=\left\langle CC^{\dagger }\right\rangle _{\mathrm{GS}}$ for bosons and
fermions respectively, where $\delta R=R-\Delta _{R}$ is the quadrature
fluctuation around its mean value, and $C=(c,c^{\dagger })^{T}$ is defined
in the Nambu space $c=(f,c_{j,\uparrow ,\mathrm{L}},c_{j,\downarrow ,\mathrm{%
L}},c_{j,\uparrow ,\mathrm{R}},c_{j,\downarrow ,\mathrm{R}})^{T}$.
Alternatively, one can introduce the Majorana basis $A$ using $A=W_{f}C$
with
\begin{equation}
W_{f}=\left(
\begin{array}{cc}
{\openone}_{4N+1} & {\openone}_{4N+1} \\
-i{\openone}_{4N+1} & i{\openone}_{4N+1}%
\end{array}%
\right) .
\end{equation}%
Majorana operators are given by linear combinations of the form $(c_{\alpha
}+c_{\alpha }^{\dagger })$, $i(c_{\alpha }^{\dagger }-c_{\alpha })$. Then,
the covariance matrix is defined as $\Gamma _{m}=i\left\langle [A,A^{\dagger
}]\right\rangle _{\mathrm{GS}}/2$ \cite{Shi2018,Kraus2010}.

In the next section we will obtain the variational ground state of the
Anderson-Holstein model by analyzing the flow of variational parameters $%
\lambda $, $\Delta _{R}$, and $\Gamma _{b,f(m)}$ in imaginary time. We will
then derive the real time equations of motion (EOM) which we will apply to
study the dynamics.

\section{Ground state properties}

In this section, we use imaginary time evolution to approximate the ground
state of the Anderson-Holstein model. We will demonstrate that in the case
of the canonical Anderson model our method correctly captures the
non-perturbative Kondo effect, including the formation of a resonance at the
Fermi energy and presence of characteristic spin correlations between the
impurity and reservoir electrons indicating the presence of Kondo spin
screening clouds. We project the equations of motion (EOM)%
\begin{equation}
\partial _{\tau }\left\vert \Psi \right\rangle =-(H-\left\langle
H\right\rangle )\left\vert \Psi \right\rangle  \label{IM}
\end{equation}%
onto the tangential plane of the variational manifold (\ref{NGS}), and
obtain the EOM for the variational parameters $\lambda $, $\Delta _{R}$, and
$\Gamma _{b,f(m)}$ (see Ref. \cite{Shi2018} for details). As $\tau
\rightarrow \infty $ the system reaches a fixed point which approximates to
the ground state of the system.

The steady state solution of the EOM in the limit $\tau \rightarrow \infty $
determines the ground state properties, including the occupation number, the
magnetization, and the spectral function of electrons in HOMO, as well as
correlation functions between HOMO and reservoirs.

\subsection{Equations of motion in imaginary time}

We first analyze the structure of the tangential vectors of the variational
manifold (\ref{NGS}). To this purpose, we introduce the unitary operator%
\begin{equation}
U_{\mathrm{GS}}=e^{-\frac{1}{2}R^{T}\sigma ^{y}\Delta _{R}}e^{-i\frac{1}{4}%
R^{T}\xi _{b}R}e^{i\frac{1}{2}C^{\dagger }\xi _{f}C}
\end{equation}%
that generates the Gaussian state and transforms $R$ and $C$ as $U_{\mathrm{%
GS}}^{\dagger }RU_{\mathrm{GS}}=SR+\Delta _{R}$ and $U_{\mathrm{GS}%
}^{\dagger }CU_{\mathrm{GS}}=U_{f}C$. The covariance matrices $\Gamma
_{b}=SS^{T}$ and $\Gamma _{f}=U_{f}(1+\sigma ^{z})U_{f}^{\dagger }/2$ are
related to the symplectic and unitary transformations via $S=e^{i\sigma
^{y}\xi _{b}}$\ and $U_{f}=e^{i\xi _{f}}$.

Let us consider the tangential vector for the variational wavefunction (\ref%
{NGS}). It is defined as the derivative of $\left\vert \Psi _{\mathrm{NGS}%
}\right\rangle $ with respect to $\tau $
\begin{equation}
\partial _{\tau }\left\vert \Psi _{\mathrm{NGS}}\right\rangle =U_{\mathrm{ph}%
}U_{\mathrm{A}}U_{\mathrm{GS}}(V_{1}+V_{2}+V_{h})\left\vert 0\right\rangle ,
\label{V1_V2_V3}
\end{equation}%
We separated the $V_1$, $V_2$, $V_h$ terms in equation (\ref{V1_V2_V3})
based on the number of creation and annihilation operators.

The $V_1$ term in (\ref{V1_V2_V3}) is linear in phonon operators
\begin{equation}
V_{1}=R^{T}S^{T}[i(1-\gamma \left\langle P_{z}f_{\gamma }^{\dagger
}f_{\gamma }\right\rangle _{\mathrm{GS}})\partial _{\tau }\lambda -\frac{1}{2%
}\sigma ^{y}\partial _{\tau }\Delta _{R}]
\end{equation}%
It is determined by the expectation value $\left\langle P_{z}f_{\gamma
}^{\dagger }f_{\gamma }\right\rangle _{\mathrm{GS}}$ in the Gaussian state.

The $V_{2}$ term in (\ref{V1_V2_V3}) is quadratic in phonon and electron
creation/annihilation operators
\begin{equation}
V_{2}=-\frac{1}{4}\text{:}R^{T}S^{T}\sigma ^{y}\partial _{\tau }SR\text{:}+%
\frac{1}{2}\text{:}C^{\dagger }U_{f}^{\dagger }(\partial _{\tau
}+O_{f})U_{f}C\text{:}  \label{V2_normal_ordered}
\end{equation}%
Note that since the operator $V_{2}$ acts on the vacuum state we should
bring it to normal ordered form. This is indicated by \textquotedblleft
::\textquotedblright\ in equation (\ref{V2_normal_ordered}). Due to the
presence of matrices performing rotations of creation/annihilation
operators, $S$ for phonons and $U_{f}$ for fermions, this normal ordering is
with respect to the instantaneous Gaussian state. The matrix $O_{f}$ is
defined from
\begin{equation}
O_{f}=-\lambda ^{T}\sigma ^{y}\partial _{\tau }\lambda \sigma ^{z}\otimes
I_{f}+\gamma (2\lambda ^{T}\sigma ^{y}-i\Delta _{R}^{T})\partial _{\tau
}\lambda O_{P}
\end{equation}%
where the diagonal matrix $I_{f}$ has only one non-zero element $%
(I_{f})_{11}=1$, and%
\begin{equation}
O_{P}=2iW_{f}^{\dagger }\frac{\delta }{\delta \Gamma _{m}}\left\langle
P_{z}f_{\gamma }^{\dagger }f_{\gamma }\right\rangle _{\mathrm{GS}}W_{f};
\end{equation}

The $V_h$ term in equation (\ref{V1_V2_V3})
\begin{eqnarray}
V_{h} &=&\gamma \lbrack (2\lambda ^{T}\sigma ^{y}-i\Delta _{R}^{T})U_{%
\mathrm{GS}}^{\dagger }P_{h}U_{\mathrm{GS}}  \notag \\
&&-iR^{T}S^{T}U_{\mathrm{GS}}^{\dagger }P_{c}U_{\mathrm{GS}}]\partial _{\tau
}\lambda
\end{eqnarray}%
contains higher order terms defined by the expansion%
\begin{eqnarray}
P_{z}f_{\gamma }^{\dagger }f_{\gamma } &=&\left\langle P_{z}f_{\gamma
}^{\dagger }f_{\gamma }\right\rangle _{\mathrm{GS}}+P_{c},  \notag \\
P_{c} &=&\frac{1}{2}\text{:}C^{\dagger }O_{P}C\text{:}+P_{h}
\label{PzFdaggerF_expansion}
\end{eqnarray}%
Equation (\ref{PzFdaggerF_expansion}) should be understood as the definition
of $P_{h}$.

The projection on the tangential vector $U_{\mathrm{ph}}U_{\mathrm{A}}U_{%
\mathrm{GS}}V_{1}\left\vert 0\right\rangle $ leads to EOM for the
expectation value of the phonon quadrature $\Delta _{R}$
\begin{eqnarray}
\partial _{\tau }\Delta _{R} &=&-\Gamma _{b}[\omega _{b}\Delta _{R}-2\lambda
\text{Im}\left\langle \mathcal{I}_{dc}\right\rangle _{\mathrm{GS}}  \notag \\
&&+2(1-\gamma \left\langle P_{z}f_{\gamma }^{\dagger }f_{\gamma
}\right\rangle _{\mathrm{GS}})G_{\lambda }]  \notag \\
&&+2i(1-\gamma \left\langle P_{z}f_{\gamma }^{\dagger }f_{\gamma
}\right\rangle _{\mathrm{GS}})\sigma ^{y}\partial _{\tau }\lambda  \label{dR}
\end{eqnarray}
where
\begin{eqnarray}
\mathcal{I}_{dc} &=&Ve_{\lambda }\sum_{\mathrm{a}}[P_{z}(f_{\gamma
}^{\dagger }+f_{\gamma })(\gamma c_{0,\uparrow ,\mathrm{a}}-c_{0,\downarrow ,%
\mathrm{a}})  \notag \\
&&+(f_{\gamma }^{\dagger }+f_{\gamma })c_{0,\uparrow ,\mathrm{a}}-\gamma
(f_{\gamma }^{\dagger }-f_{\gamma })c_{0,\downarrow ,\mathrm{a}}],
\end{eqnarray}%
and the effective hybridization strength $V_{\mathrm{eff}}=Ve_{\lambda }$ is
reduced by the factor $e_{\lambda }\equiv \left\langle e^{iR^{T}\lambda
}\right\rangle _{\mathrm{GS}}=e^{i\Delta _{R}^{T}\lambda }e^{-\frac{1}{2}%
\lambda ^{T}\Gamma _{b}\lambda }$.

The projection on the tangential vector $U_{\mathrm{ph}}U_{\mathrm{A}}U_{%
\mathrm{GS}}V_{2}\left\vert 0\right\rangle $ results in EOM%
\begin{eqnarray}
\partial _{\tau }\Gamma _{b} &=&\sigma ^{y}\Omega _{\mathrm{re}}\sigma
^{y}-\Gamma _{b}\Omega _{\mathrm{re}}\Gamma _{b},  \label{dGb} \\
\partial _{\tau }\Gamma _{m} &=&-\mathcal{H}_{m}-\Gamma _{m}\mathcal{H}%
_{m}\Gamma _{m}+i[\Gamma _{m},O_{m}]  \label{dGm}
\end{eqnarray}%
for the covariance matrices $\Gamma _{b,m}$, where $\Omega _{\mathrm{re}%
}=\omega _{b}I_{2}-2$Re$\left\langle \mathcal{I}_{dc}\right\rangle _{\mathrm{%
GS}}\lambda \lambda ^{T}$ and $\mathcal{H}_{m}$ are the mean-field single
particle Hamiltonians of phonon and electrons, and $%
O_{m}=-iW_{f}O_{f}W_{f}^{\dagger }/2$. The explicit form of $\mathcal{H}_{m}$
is shown in Appendix A.

By projecting on the tangential vector $U_{\mathrm{ph}}U_{\mathrm{A}}U_{%
\mathrm{GS}}V_{h}\left\vert 0\right\rangle $, we obtain EOM%
\begin{equation}
\partial _{\tau }\lambda ^{T}\mathbf{G}\partial _{\tau }\lambda =\partial
_{\tau }\lambda ^{T}\xi _{\tau }.  \label{dL}
\end{equation}%
The explicit form of the Gram matrix $\mathbf{G}$ and the vector $\xi _{\tau
}$ is given in Appendix A. By solving Eqs. (\ref{dR}), (\ref{dGb}), (\ref%
{dGm}) and (\ref{dL}) numerically, we obtain the ground state configuration
in the limit $\tau \rightarrow \infty $. We note that the EOM (\ref{dL}) is
quadratic in $\partial _{\tau }\lambda $, but it can be reduced to the
linear ordinary differential equation (ODE) (see Appendix B).

\subsection{Physical observables}

In this subsection we will discuss several physical quantities that can be
used to characterize the ground state of the Anderson-Holstein model (here
the bias voltage $V_e=0$).

In the sector $\gamma $, the ground state energy $E_{\gamma }=E_{\mathrm{res}%
}+E_{\mathrm{mol}}+E_{V}$ is composed of a reservoir part $E_{\mathrm{res}%
}=\sum_{ij,\sigma ,\mathrm{a}}h_{\mathrm{a},ij}\left\langle c_{i,\sigma ,%
\mathrm{a}}^{\dagger }c_{j,\sigma ,\mathrm{a}}\right\rangle _{\mathrm{GS}}$,
the molecular energy $E_{\mathrm{mol}}=E_{\mathrm{A}}+E_{\mathrm{phon}}$:%
\begin{eqnarray}
E_{\mathrm{A}} &=&\tilde{\varepsilon}_{d}+\frac{1}{2}\tilde{U}\left\langle
f_{\gamma }^{\dagger }f_{\gamma }\right\rangle _{\mathrm{GS}}+\frac{1}{2}%
U_{P}\gamma \left\langle P_{z}f_{\gamma }^{\dagger }f_{\gamma }\right\rangle
_{\mathrm{GS}},  \notag \\
E_{\mathrm{phon}} &=&\frac{1}{4}\omega _{b}(\Delta _{R}^{T}\Delta
_{R}+tr\Gamma _{b})-\frac{1}{2}\omega _{b}+\Delta _{R}^{T}G_{\lambda }
\notag \\
&&+4(\omega _{b}\lambda ^{T}\lambda -2g\lambda _{p}),
\end{eqnarray}%
and the hybridization energy $E_{V}=$Re$\left\langle \mathcal{I}%
_{dc}\right\rangle _{\mathrm{GS}}$, where $h_{\mathrm{a}=\mathrm{L,R}%
}=I_{2}\otimes (-t_{0}\delta _{i,j\pm 1}-\mu _{\mathrm{a}}\delta _{ij})$. As
pointed out in Ref \cite{Shi2018} the energy $E_{\gamma }$ monotonically
decreases during the imaginary time evolution.

The occupation number and the magnetization of electrons in HOMO are given
by $\left\langle n_{d}\right\rangle =1+\gamma \left\langle P_{z}f_{\gamma
}^{\dagger }f_{\gamma }\right\rangle _{\mathrm{GS}}$ and $m_{z}=\gamma
\left\langle P_{z}(1-f_{\gamma }^{\dagger }f_{\gamma })\right\rangle _{%
\mathrm{GS}}$, respectively. Correlation between electrons in the HOMO and
reservoirs are characterized by the correlation functions $C_{\alpha
=x,y,z}=\left\langle d^{\dagger }\tau ^{\alpha }dc_{j}^{\dagger }\tau
^{\alpha }c_{j}\right\rangle /4$, where $\tau ^{\alpha }$ is the Pauli
matrix. In the transformed frame, the correlation functions can be expressed
as expectation values computed in the Gaussian state
\begin{eqnarray}
C_{x} &=&\frac{1}{4}\gamma \left\langle (1-f_{\gamma }^{\dagger }f_{\gamma
})c_{j}^{\dagger }\tau ^{x}c_{j}\right\rangle _{\mathrm{GS}},  \notag \\
C_{y} &=&-i\frac{1}{4}\left\langle P_{z}(1-f_{\gamma }^{\dagger }f_{\gamma
})c_{j}^{\dagger }\tau ^{y}c_{j}\right\rangle _{\mathrm{GS}},  \notag \\
C_{z} &=&\frac{1}{4}\gamma \left\langle P_{z}(1-f_{\gamma }^{\dagger
}f_{\gamma })c_{j}^{\dagger }\tau ^{z}c_{j}\right\rangle _{\mathrm{GS}}
\end{eqnarray}%
The hole excitations in HOMO interact with the charge in the substrate via
the Coulomb interaction, which results in the displacement%
\begin{equation}
\left\langle R\right\rangle =\Delta _{R}-2i\sigma ^{y}\lambda (1-\gamma
\left\langle P_{z}f_{\gamma }^{\dagger }f_{\gamma }\right\rangle _{\mathrm{GS%
}})
\end{equation}%
of phonon.

The spectral function $\mathcal{A}(\omega )=-$Im$G_{R}(\omega )/\pi $
characterizes the properties of excitations above the ground state. More
specifically, for the Anderson-Holstein system, it is determined by the
Fourier transform $G_{R}(\omega )=\int dte^{i\omega t}G_{R}(t)$ of the
retarded Green function%
\begin{equation}
G_{R}(t)=-i\left\langle \{d_{\downarrow }(t),d_{\downarrow }^{\dagger
}(0)\}\right\rangle \theta (t).
\end{equation}%
Following the unitary transformation given by $U_{\mathrm{ph}}U_{\mathrm{A}}$%
, the Green's function becomes%
\begin{equation}
G_{R}(t)=-i\left\langle \{\bar{d}_{\downarrow }(t),\bar{d}_{\downarrow
}^{\dagger }(0)\}\right\rangle _{\mathrm{GS}}\theta (t),
\end{equation}%
where the evolution $\bar{d}_{\downarrow }(t)=e^{iH_{2}t}\bar{d}_{\downarrow
}e^{-iH_{2}t}$ of the fermionic operator $\bar{d}_{\downarrow
}=e^{-iR^{T}\lambda }F$ is governed by the Hamiltonian $H_{2}$, and%
\begin{equation}
F=\frac{1}{2}[\gamma (f_{\gamma }^{\dagger }-f_{\gamma })-P_{z}(f_{\gamma
}^{\dagger }+f_{\gamma })].
\end{equation}

We approximate $\bar{d}_{\downarrow }(t)\sim e^{iH_{\mathrm{MF}}t}\bar{d}%
_{\downarrow }e^{-iH_{\mathrm{MF}}t}$ by the mean-field Hamiltonian $H_{%
\mathrm{MF}}=H_{\mathrm{MF}}^{p}+H_{\mathrm{MF}}^{e}$ with $H_{\mathrm{MF}%
}^{p}=\delta R^{T}\Omega _{\mathrm{re}}\delta R/4$ and $H_{\mathrm{MF}%
}^{e}=iA^{T}\mathcal{H}_{m}A/4$, where $\Omega _{\mathrm{re}}$ and $\mathcal{%
H}_{m}$ are determined by the average value of quadrature and covariance
matrices in the ground state. Since the bosonic and electronic parts in the
Gaussian state are factorized, the retarded Green function reads%
\begin{eqnarray}
G_{R}(t) &=&-i\theta (t)[\left\langle e^{-iR^{T}(t)\lambda }e^{iR^{T}\lambda
}\right\rangle _{\mathrm{GS}}\left\langle F(t)F^{\dagger }\right\rangle _{%
\mathrm{GS}}  \notag \\
&&+\left\langle e^{iR^{T}\lambda }e^{-iR^{T}(t)\lambda }\right\rangle _{%
\mathrm{GS}}\left\langle F^{\dagger }F(t)\right\rangle _{\mathrm{GS}}].
\end{eqnarray}

The Green function $G_{R}(t)$ contains the average values on the Gaussian
state, which can be obtained analytically in terms of the covariance
matrices and $\Delta _{R}$ (see Appendix C). For instance,%
\begin{eqnarray}
\left\langle e^{-iR^{T}(t)\lambda }e^{iR^{T}\lambda }\right\rangle _{\mathrm{%
GS}} &=&e^{-\alpha }\sum_{n=0}^{\infty }\frac{\alpha ^{n}}{n!}e^{-in\omega _{%
\mathrm{re}}t},  \notag \\
\left\langle e^{iR^{T}\lambda }e^{-iR^{T}(t)\lambda }\right\rangle _{\mathrm{%
GS}} &=&e^{-\alpha }\sum_{n=0}^{\infty }\frac{\alpha ^{n}}{n!}e^{in\omega _{%
\mathrm{re}}t}.
\end{eqnarray}%
where $\alpha =\lambda ^{T}\Gamma _{b}\lambda $ and $\omega _{\mathrm{re}}$
is the simplectic eigenvalue of $\Omega _{\mathrm{re}}$.

Eventually, the imaginary part of the Fourier transform%
\begin{eqnarray}
G^{>}(\omega ) &=&-i\int_{0}^{\infty }dte^{i(\omega +i\delta )t}\left\langle
F(t)F^{\dagger }\right\rangle _{\mathrm{GS}},  \notag \\
G^{<}(\omega ) &=&-i\int_{0}^{\infty }dte^{i(\omega +i\delta )t}\left\langle
F^{\dagger }F(t)\right\rangle _{\mathrm{GS}}
\end{eqnarray}%
gives the spectral function%
\begin{eqnarray}
\mathcal{A}(\omega ) &=&-\frac{1}{\pi }e^{-\alpha }\sum_{n=0}^{\infty }\frac{%
\alpha ^{n}}{n!}\text{Im}[G^{>}(\omega -n\omega _{\mathrm{re}})  \notag \\
&&+G^{<}(\omega +n\omega _{\mathrm{re}})],  \label{Sp}
\end{eqnarray}%
where we add a small imaginary part $i\delta $ in the numerical calculation
of Fourier transforms $G^{>(<)}$.

In the next two subsections, we show the occupation number, the correlation
functions, the displacement, and the spectral function for the
Anderson-Holstein model.

\subsection{Anderson model in equilibrium}

In this subsection, we present results for the ground state properties of
the Anderson model, i.e. when the electron-phonon coupling $g=0$. We expect
that in the Kondo regime $\varepsilon _{d}<0$, $\varepsilon _{d}+U>0$, and $%
\Gamma =V^{2}<(U,\varepsilon _{d})$, HOMO is singly occupied, i.e., $%
\left\langle n_{d}\right\rangle \sim 1$ and its magnetization $m_{z}=0$,
which reflects screening of the impurity spin by electrons in reservoirs. An
important signature of the Kondo regime of the Anderson model is the
formation of anti-ferromagnetic correlations between spins of electrons in
HOMO and on the adjacent sites in the reservoirs. These correlations will be
absent when the impurity is in the doubly occupied regime with $\varepsilon
_{d}<\varepsilon _{d}+U<0$ and $\left\langle n_{d}\right\rangle \sim 2$,
which is relevant to the HOMO in the THz STM experiments.

\begin{figure}[tbp]
\includegraphics[width=0.99\linewidth]{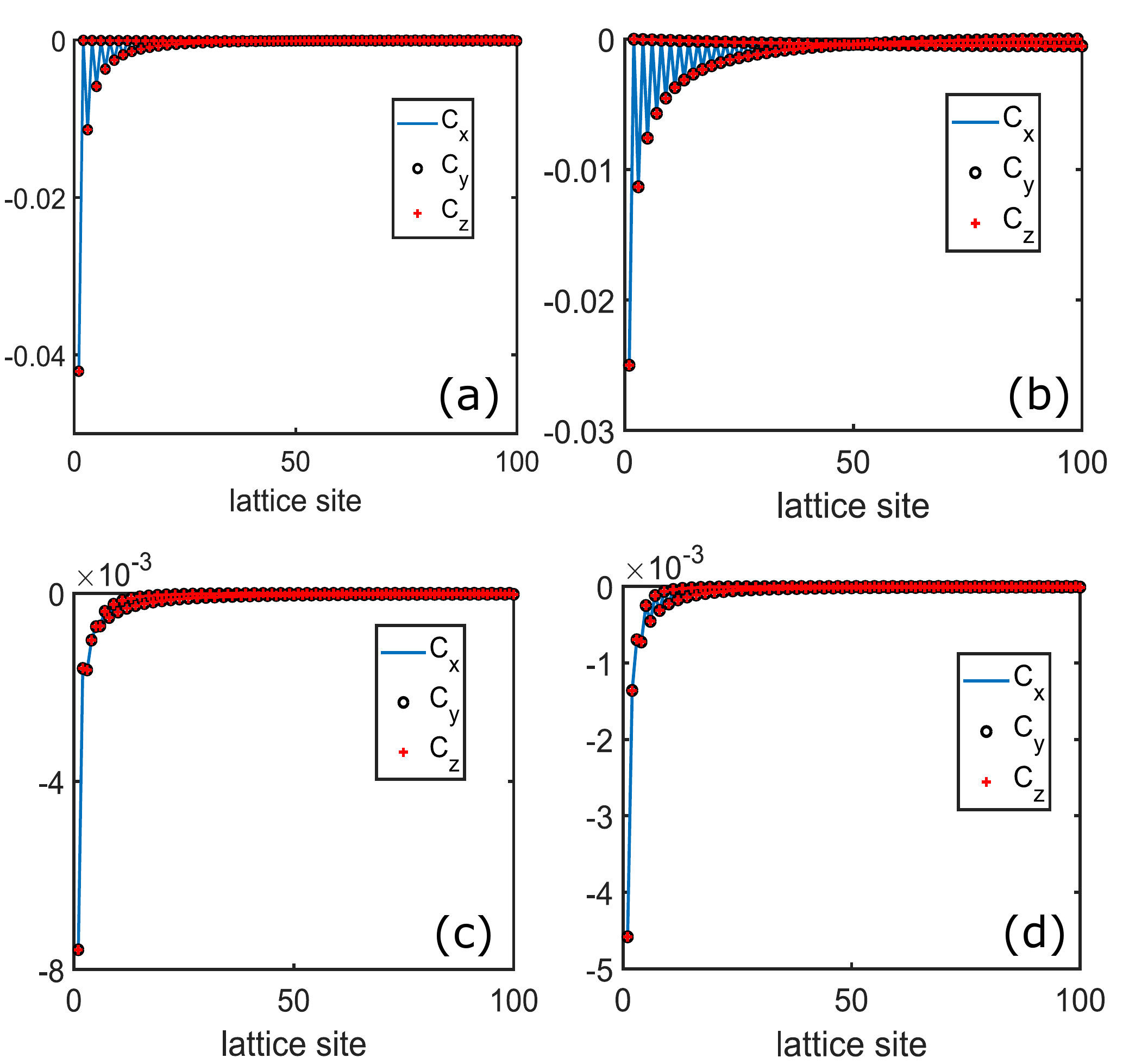}
\caption{The correlation functions $C_{\protect\alpha}$ in the Kondo regime
(a)-(b) and the double occupation regime (c)-(d), where $\protect\varepsilon%
_d=-0.5$ the hopping amplitude $t_0$ is taken as the unit. (a) $U=1$ and $%
\Gamma=0.16$; (b) $U=1$ and $\Gamma=0.04$; (c) $U=0.25$ and $\Gamma=0.04$;
(c) $U=0.05$ and $\Gamma=0.04$.}
\label{CFGSwithoutphonon}
\end{figure}

In Fig. \ref{CFGSwithoutphonon}, we plot the correlation functions $%
C_{\alpha =x,y,z}$ of the electrons in HOMO and one of the reservoirs in the
Kondo regime $(U,\varepsilon _{d},\Gamma )=(1,-0.5,0.16)$ and $(1,-0.5,0.04)$%
, and the double occupation regime $(0.25,-0.5,0.04)$ and $(0.05,-0.5,0.04)$%
. In the Kondo regime (Figs. \ref{CFGSwithoutphonon}a and \ref%
{CFGSwithoutphonon}b), the $SU(2)$-symmetric $C_{\alpha }$ shows significant
anti-ferromagnetic correlations of electrons in the HOMO and the reservoirs.
In the double occupation regime (Figs. \ref{CFGSwithoutphonon}c and \ref%
{CFGSwithoutphonon}d) we observe small values of the correlations $C_{\alpha
}\sim 10^{-3}$, which indicates weak correlations between the singlet
electron pair in the HOMO and the reservoirs.

\begin{figure}[tbp]
\includegraphics[width=0.99\linewidth]{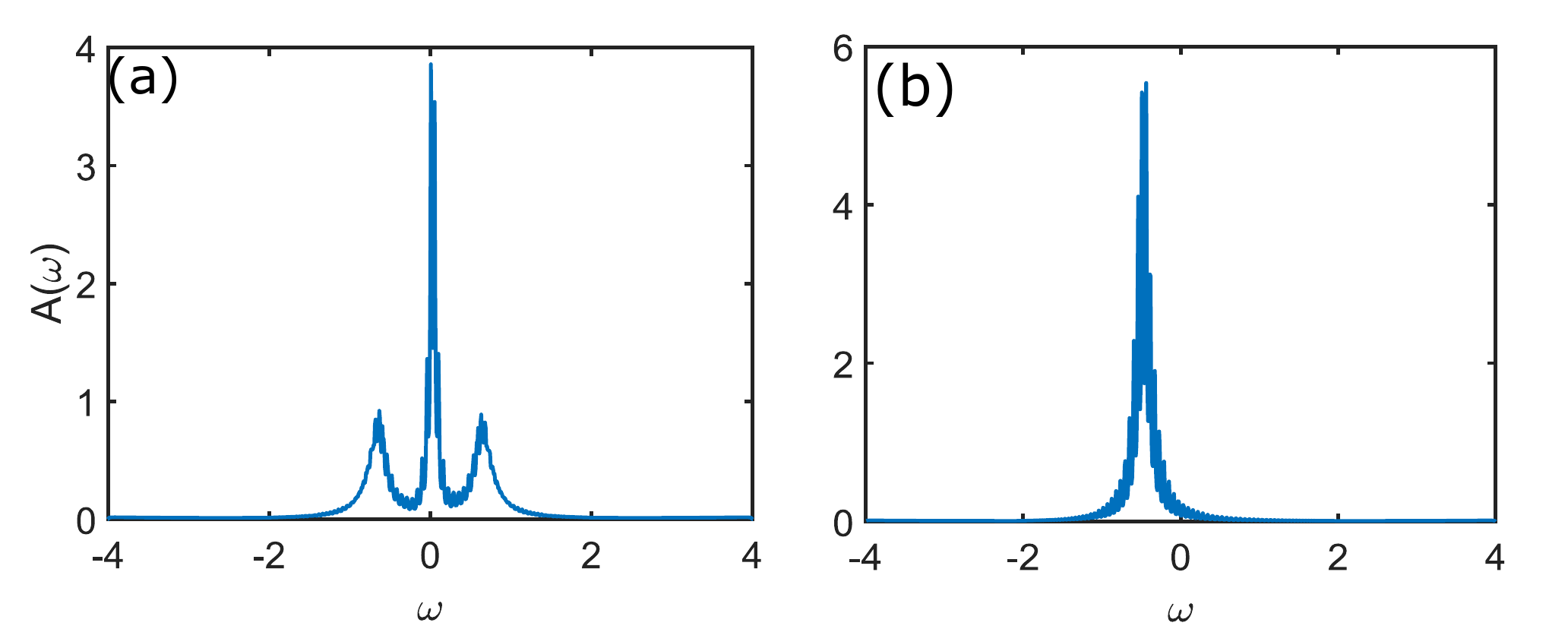}
\caption{Electron spectral function in the localized orbital in the Kondo
and double occupancy regimes without the electron-phonon interaction. System
size is 100 and the hopping amplitude $t_0$ is taken as the unit: (a) $(U,%
\protect\varepsilon _{d},\Gamma )=(1,-0.5,0.04)$ in the Kondo regime; (b) $%
(U,\protect\varepsilon _{d},\Gamma )=(0.05,-0.5,0.04)$ in the double
occupancy regime.}
\label{SFwithoutphonon}
\end{figure}

In Fig. \ref{SFwithoutphonon}, we show the spectral function $\mathcal{A}
(\omega )$ in the Kondo and the doubly occupied regimes, where we take $%
\delta =0.01$. In Fig. \ref{SFwithoutphonon}a, the spectral function in the
Kondo regime $(U,\varepsilon _{d},\Gamma )=(1,-0.5,0.04)$ displays both the
Kondo resonance peak around $\omega =0$ and the resonance with energy levels
$\varepsilon _{d}$ and $\varepsilon _{d}+U$. In the the double occupation
regime $(U,\varepsilon _{d},\Gamma )=(0.05,-0.5,0.04)$, as shown in Fig. \ref%
{SFwithoutphonon}b, the Kondo resonance peak vanishes.

\subsection{Anderson-Holstein model}

In this subsection, we present results for the equilibrium properties of the
full Anderson-Holstein model with finite electron-phonon interaction.
Consequencies of the electron-phonon interaction $g$ include a change of the
effective single particle energy level, softening of the on-site repulsive
interaction, and polaronic dressing of hybridization. Note, for example,
that when the occupation number on the molecular orbital $n_{d}$ is
different from two, we get a finite displacement of the phonon operator $%
x_{0}=\left\langle x\right\rangle $, which favors partial occupation of the
HOMO. We observe that in the Kondo regime including the electron-phonon
intraction tends to suppress the formation of the singlet cloud, while in
the doubly occupied regime it favors the creation of the hole excitations in
HOMO.

\begin{figure}[tbp]
\includegraphics[width=0.99\linewidth]{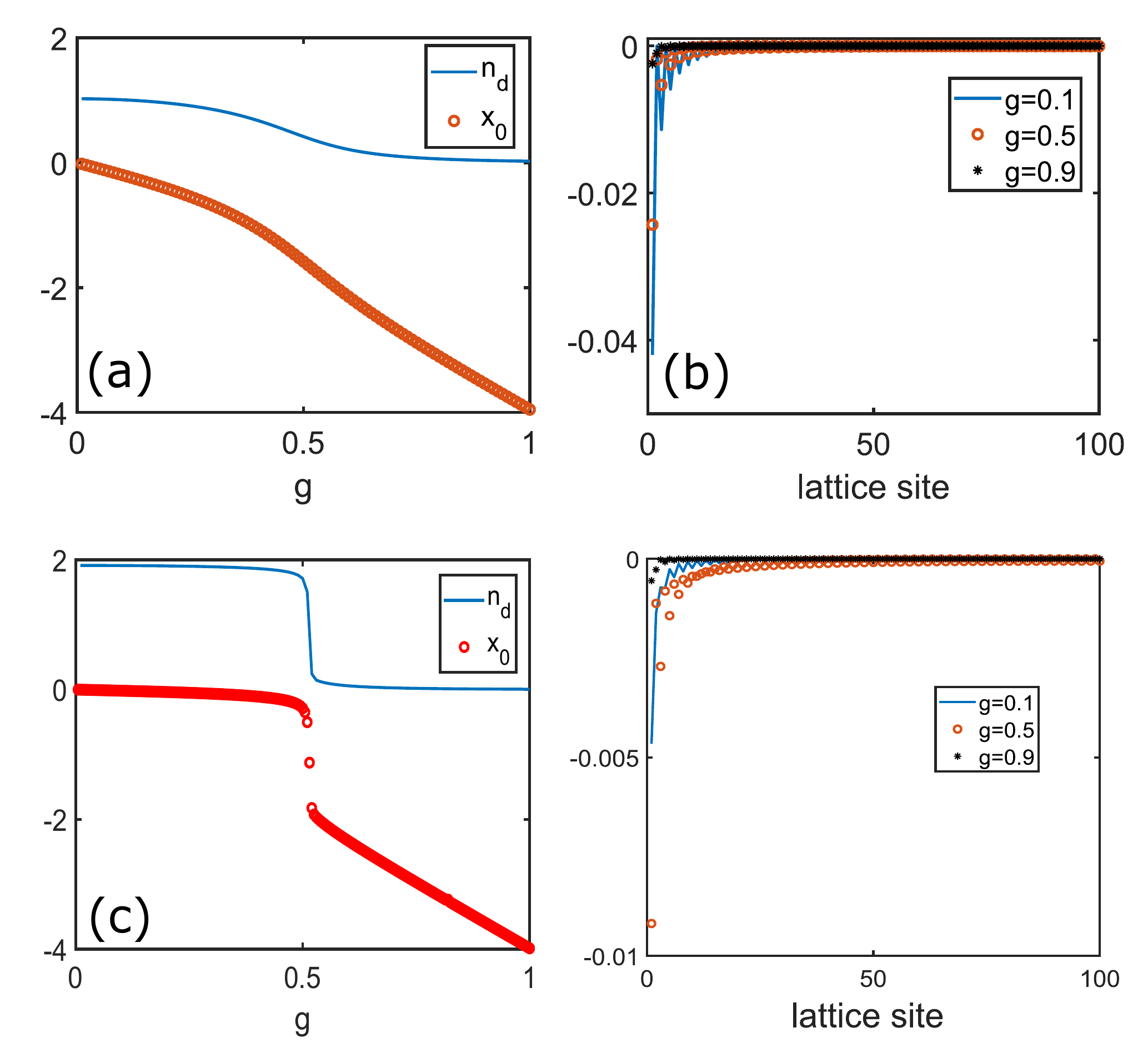}
\caption{The occupation number $n_d$, the displacement $x_0$, and the
correlation function $C_{\protect\alpha}$ in the Kondo regime (a)-(b) and
the double occupation regime (c)-(d), where $\protect\varepsilon_d=-0.5$, $%
\protect\omega _{b}=1$, and the hopping amplitude $t_0$ is taken as the
unit. (a) The occupation number $n_d$ and the displacement $x_0$ for $U=1$
and $\Gamma=0.16$; (b) The correlation function $C_{\protect\alpha}$ for $%
U=1 $ and $\Gamma=0.16$; (c) The occupation number $n_d$ and the
displacement $x_0$ for $U=0.05$ and $\Gamma=0.04$; (d) The correlation
function $C_{\protect\alpha}$ for $U=0.05$ and $\Gamma=0.04$.}
\label{withphonon}
\end{figure}

In Fig. \ref{withphonon}a, we show the occupation number $n_{d}$ and the
displacement $x_{0}$ as a function of the coupling strength $g$ in the Kondo
regime $(U,\varepsilon _{d},\Gamma )=(1,-0.5,0.16)$, where the magnetization
$m_{z}=0$. In Fig. \ref{withphonon}b, we show the correlation function $%
C_{\alpha }$ for different $g=0.1$, $0.5$, and $0.9$ in the Kondo regime. As
$g$ increases, the displacement $x_{0}$ becomes larger, which lifts the
single particle energy level of HOMO, therefore the the occupation number $%
n_{d}$ decreases from $1$ to $0$. The finite electron-phonon interaction
also softens the on-site interaction and the effective hybridization
strength $V_{\mathrm{eff}}<V$, thus, as shown in Fig. \ref{withphonon}b, the
larger the coupling strength $g$, the weaker the correlation $C_{\alpha }$
between HOMO and leads. Eventually, the Kondo singlet is destroyed by the
finite electron-phonon coupling.

In Fig. \ref{withphonon}c, we show the occupation number $n_{d}$ and the
displacement $x_{0}$ as a function of $g$ in the double occupation regime $%
(U,\varepsilon _{d},\Gamma )=(0.05,-0.5,0.04)$, where the magnetization $%
m_{z}=0$. In Fig. \ref{withphonon}d, we show the correlation functions $%
C_{\alpha }$ for different $g=0.1$, $0.5$, and $0.9$ in the double
occupation regime. Since the single particle energy level is lifted, the
particle number $n_{d}$ decreases from $2$ to $0$ as $g$ increases. For
small electron-phonon couplings, e.g., $g=0.1$, HOMO is mostly doubly
occupied, and $C_{\alpha }$ displays a weak correlation of HOMO and
reservoirs. In the intermediate coupling regime, e.g., $g=0.5$, the
occupation number $n_{d}$ decreases to $1.5$, meaning that electrons in HOMO
are in the superposition of doubly and singly occupied states. Since the
total spin in HOMO is totally screened, i.e., $m_{z}=0$, the singly occupied
electron forms the singlet state with lead electrons. This explains somewhat
the presence of a stronger correlation $C_{\alpha }$ in the intermediate
regime than that in the weakly coupling regime. When $g$ increases further,
e.g. to $g=0.9$, the occupation number $n_{d}$ is reduced to $0$, and the
correlation $C_{\alpha }$ becomes vanishingly small.

\begin{figure}[tbp]
\includegraphics[width=0.99\linewidth]{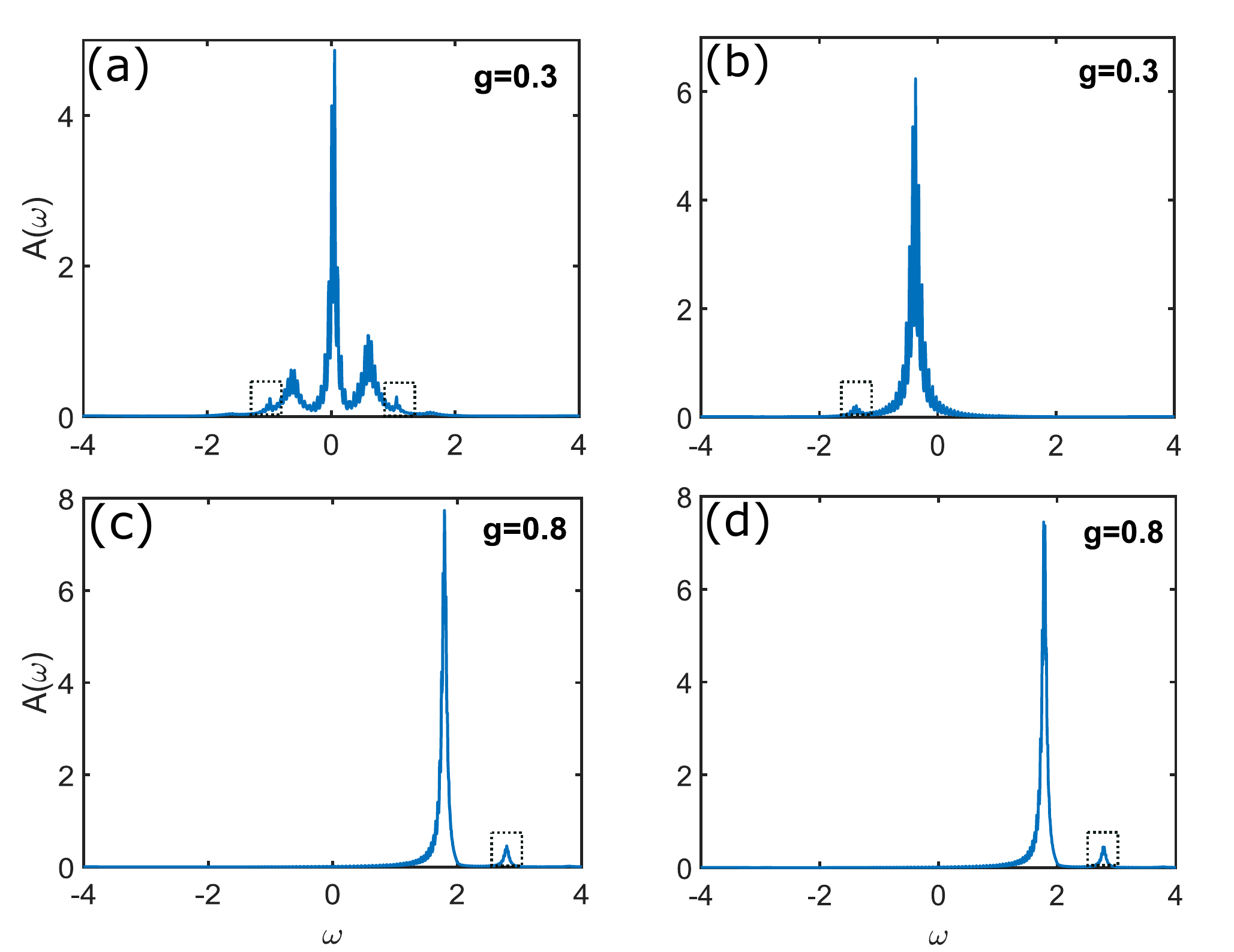}
\caption{Electron spectral function in the localized orbital in the Kondo
and double occupancy regimes including electron-phonon interaction. System
size is 100 and the hopping amplitude $t_0$ is taken as the unit: (a) and
(c) $(U,\protect\varepsilon _{d},\Gamma )=(1,-0.5,0.04)$ in the Kondo
regime; (b) and (d) $(U,\protect\varepsilon _{d},\Gamma )=(0.05,-0.5,0.04)$
in the double occupancy regime.}
\label{SFwithphonon}
\end{figure}

In Fig. \ref{SFwithphonon}, we show the spectral function $\mathcal{A}%
(\omega )$ in the Kondo and doubly occupied regimes. In Fig. \ref%
{SFwithphonon}a, the spectral function $\mathcal{A}(\omega )$ in the Kondo
regime $(U,\varepsilon _{d},\Gamma )=(1,-0.5,0.04)$ is shown, where the
coupling strength $g=0.2$ and $0.4$. For $g=0.2$, the renormalized single
particle energy level is slightly lifted to $\tilde{\varepsilon}_{d}\sim
-0.40$ and the on-site interaction $\tilde{U}\sim 0.93$ is softened due to
the electron-phonon interaction. Since $\tilde{\varepsilon}_{d}<0$ and $%
\tilde{\varepsilon}_{d}+\tilde{U}>0$ still in the Kondo regime, the Kondo
peak survives in the spectral function. However, the system deviates from
the symmetric Anderson model, i.e., $\left\vert \tilde{\varepsilon}%
_{d}\right\vert \neq \tilde{\varepsilon}_{d}+\tilde{U}$, therefore, in the
spectral function, the two peaks around $\tilde{\varepsilon}_{d}$ and $%
\tilde{\varepsilon}_{d}+\tilde{U}$ become asymmetric. The small $\alpha \sim
0.02$ shows that the electron excitation in HOMO is weakly dressed by the
phonon, and the spectral function is dominated by the pure electron
excitation with the phonon in the vacuum state, i.e., $n=0$ in Eq. (\ref{Sp}%
). For the larger $g=0.4$, $\tilde{\varepsilon}_{d}$ is shifted to $-0.06$
and the on-site interaction is reduced to $\tilde{U}\sim 0.71$. The spectral
function shows that the Kondo peak is destroyed, and two peaks around $%
\tilde{\varepsilon}_{d}$ and $\tilde{\varepsilon}_{d}+\tilde{U}$ still
survive. Since the electron excitation in HOMO is dressed by the phonon with
larger average phonon number $\alpha \sim 0.08$, two peaks corresponding to
the single phonon excitation appear in the spectral function, as shown in
the (black) box of Fig. \ref{SFwithphonon}a.

In Fig. \ref{SFwithphonon}b we show the spectral function $\mathcal{A}%
(\omega )$ in the doubly occupied regime $(U,\varepsilon _{d},\Gamma
)=(0.05,-0.5,0.04)$ for two values of the electron-phonon coupling strength $%
g=0.2$ and $0.4$. We point out that for $g=0.2$ the renormalized single
particle energy level is shifted to $\tilde{\varepsilon}_{d}\sim -0.41$ and
the on-site electron-electron interaction becomes effectively attractive $%
\tilde{U}\sim -0.01$ due to the phonon mediated attraction. The small $%
\alpha \sim 0.01$ implies that the electron in HOMO is weakly dressed by the
phonon mode, as a result, the contribution of the phonon excitation to the
spectral function is negligible. Comparing with the spectral function
without electron-phonon interaction, we find that the peak is shifted to the
larger frequency around the lifted single particle level $\tilde{\varepsilon}%
_{d}$. When the coupling constant is further increasing, e.g., $g=0.4$, $%
\tilde{\varepsilon}_{d}$ is shifted to $-0.1$ and the attractive interaction
$\tilde{U}\sim -0.22$ becomes stronger. The electron excitation in HOMO is
dressed by phonons with larger average number $\alpha \sim 0.06$, which
results in the visible peak corresponding to the single phonon exitation, as
shown in the (black) box of Fig. \ref{SFwithphonon}b.

\section{Real time dynamics}

In this section, we apply the non-Gaussian ansatz (\ref{NGS}) to study
real-time dynamics for the Anderson-Holstein system. Similar to the
procedure used in the imaginary time evolution, we project the Schr\"{o}%
dinger equation%
\begin{equation}
i\partial _{t}\left\vert \Psi \right\rangle =H\left\vert \Psi \right\rangle
\label{SE}
\end{equation}%
to the tangential space of the variational manifold (\ref{NGS}) and obtain
EOM for variational parameters $\lambda $, $\Delta _{R}$, and $\Gamma
_{b,f(m)}$. Analysis of these differential equations allows us to explore a
large variety of non-equilibrium phenomena

We now present results for transport through the molecule in the two cases:
with and without electron-phonon interaction. To find DC conductance we
perform a quench-type protocol, when we start with the reservoirs at
different chemical potentials but disconnected from the molecule, and hence
from each other. We switch on the molecule-reservoirs coupling and then
evolve the system in real time until it reaches the steady state but before
electron wavepackets reflected from the outer ends arrive back at the
molecule. The steady state current $I$ as a function of (finite) $V_{e}$
gives us the full nonlinear current voltage characteristic of the system.
Differential conductance is defined through the relation $\sigma
_{c}=\partial _{V_{e}}I$. In the case of ultrafast THz STM experiments, bias
voltage is applied as a time dependent pulse given in equation (\ref{Ve}).
This voltage pulse results in a transfer of a finite number of electrons
between the two reservoirs. We compute the full time dependent evolution of
the current in the system. We demonstrate that in the case of finite
electron-phonon interaction a burst of current gives rise to vibrations of
the molecules which persist well after the duration of the pulse. This was
observed in experiments by Cocker et al. \cite{Cocker2016}.

\subsection{Equations of motion}

Projection of Eq. (\ref{SE}) to the tangential vectors $U_{\mathrm{ph}}U_{%
\mathrm{A}}U_{\mathrm{GS}}V_{1,2,h}\left\vert 0\right\rangle $ gives EOM for
variational parameters in the real time evolution. When we perform
projection on the vectors $U_{\mathrm{ph}}U_{\mathrm{A}}U_{\mathrm{GS}%
}V_{1,2}\left\vert 0\right\rangle $ we find in%
\begin{eqnarray}
\partial _{t}\Delta _{R} &=&i\sigma ^{y}[\omega _{b}\Delta _{R}-2\lambda
\text{Im}\left\langle \mathcal{I}_{dc}\right\rangle _{\mathrm{GS}}  \notag \\
&&+2(1-\gamma \left\langle P_{z}f_{\gamma }^{\dagger }f_{\gamma
}\right\rangle _{\mathrm{GS}})G_{\lambda }  \notag \\
&&+2(1-\gamma \left\langle P_{z}f_{\gamma }^{\dagger }f_{\gamma
}\right\rangle _{\mathrm{GS}})\partial _{t}\lambda ],  \label{dDt}
\end{eqnarray}%
\begin{equation}
\partial _{t}\Gamma _{b}=i(\sigma ^{y}\Omega _{\mathrm{re}}\Gamma
_{b}-\Gamma _{b}\Omega _{\mathrm{re}}\sigma ^{y}),  \label{dGbt}
\end{equation}%
and%
\begin{equation}
\partial _{t}\Gamma _{m}=[\mathcal{H}_{m}-iO_{m},\Gamma _{m}].  \label{dGmt}
\end{equation}%
To obtain the time evolution of $\lambda $, we perform the projection on the
tangential vector $U_{\mathrm{ph}}U_{\mathrm{A}}U_{\mathrm{GS}%
}V_{h}\left\vert 0\right\rangle $ and obtain
\begin{equation}
\partial _{t}\lambda ^{T}\mathbf{M}\partial _{t}\lambda =\partial
_{t}\lambda ^{T}\xi _{t}  \label{dLt}
\end{equation}%
where $\xi _{t}=i\xi _{\tau }$. While Eq. (\ref{dLt}) is non-linear in $%
\lambda $, it can be reduced to a linear ODE. Details are presented in
Appendix B.

Thus, the description of a broad range of non-equilibrium phenomena in the
Anderson-Holstein model can be reduced to solving Eqs. (\ref{dDt})-(\ref{dLt}%
) for the time evolution of variational parameters.

Before presenting results of our analysis we comment on one important
technical aspect of the calculations. The total electron number operators%
\begin{eqnarray}
N_{\uparrow } &=&\frac{1}{2}(1+\gamma P_{z})+\sum_{i,\mathrm{a}%
}c_{i,\uparrow ,\mathrm{a}}^{\dagger }c_{i,\uparrow ,\mathrm{a}},  \notag \\
N_{\downarrow } &=&\frac{1}{2}(1-\gamma P_{z})+\gamma P_{z}f_{\gamma
}^{\dagger }f_{\gamma }+\sum_{i,\mathrm{a}}c_{i,\downarrow ,\mathrm{a}%
}^{\dagger }c_{i,\downarrow ,\mathrm{a}}
\end{eqnarray}%
should be conserved throughout the real time evolution. This fundamental
conservation law should not be affected by the transformations of the
Hamiltonian $U_{ph}$ and $U_{A}$. However, when computing long time
evolution needed for finding steady states the numerical errors may
accumulate and lead to the violation of particle number conservation. To
circumvent this numerical problem, we introduce a penalty term%
\begin{equation}
H_{\Lambda }=\Lambda \lbrack (N_{\uparrow }-\bar{N}_{\uparrow
})^{2}+(N_{\downarrow }-\bar{N}_{\downarrow })^{2}]
\end{equation}%
in the Hamiltonian, where $\Lambda $ is chosen to be much larger than all
the energy scales in the system, and $\bar{N}_{\uparrow (\downarrow )}$ is
the average number of spin-up (down) electrons. Note that specific value of $%
\Lambda $ turns out to be unimportant for all results that we discuss in
this paper. The mean-field Hamiltonian $\mathcal{H}_{\Lambda ,m}$ in the
Majorana basis for the penalty term $H_{\Lambda }$ is derived in Appendix D,
which modifies the mean-field Hamiltonian (\ref{Hm}) as $\mathcal{H}%
_{m}\rightarrow \mathcal{\bar{H}}_{m}=\mathcal{H}_{m}+\mathcal{H}_{\Lambda
,m}$. In Eq. (\ref{dGmt}), the substitution of $\mathcal{H}_{m}$ by $%
\mathcal{\bar{H}}_{m}$ leads to the conserved particle numbers $\left\langle
N_{\uparrow (\downarrow )}\right\rangle =N_{\uparrow (\downarrow )}^{0}$ in
the real time evolution.

\begin{figure}[tbp]
\includegraphics[width=0.99\linewidth]{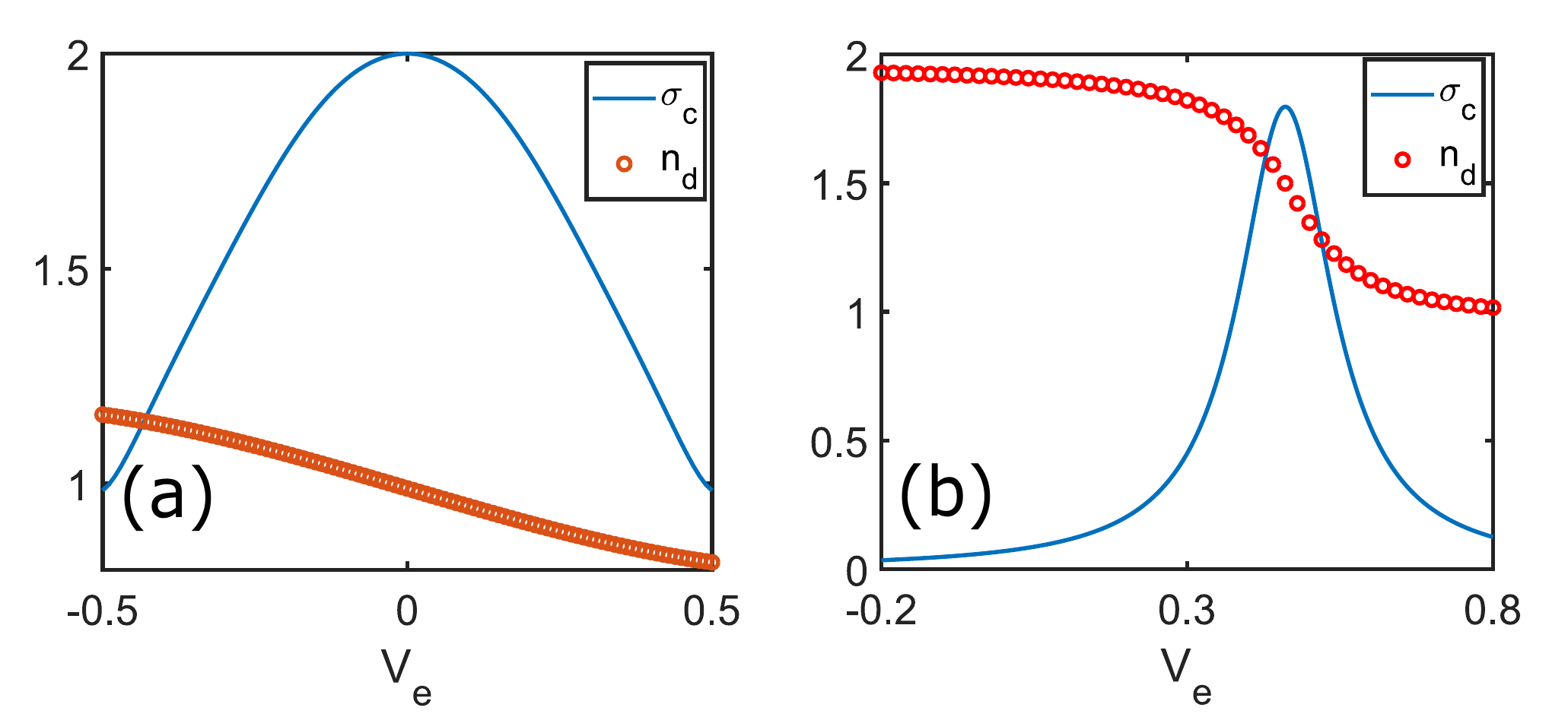}
\caption{The occupation number $n_d$ and the finite bias conductance $%
\protect\sigma_c$ in the Kondo regime (a) $U=1$ and $\Gamma=0.16$ and the
double occupation regime (b) $U=0.05$ and $\Gamma=0.04$, where $\protect%
\varepsilon_d=-0.5$ and the hopping amplitude $t_0$ is taken as the unit.}
\label{tdwithoutphonon}
\end{figure}

\begin{figure}[tbp]
\includegraphics[width=0.9\linewidth]{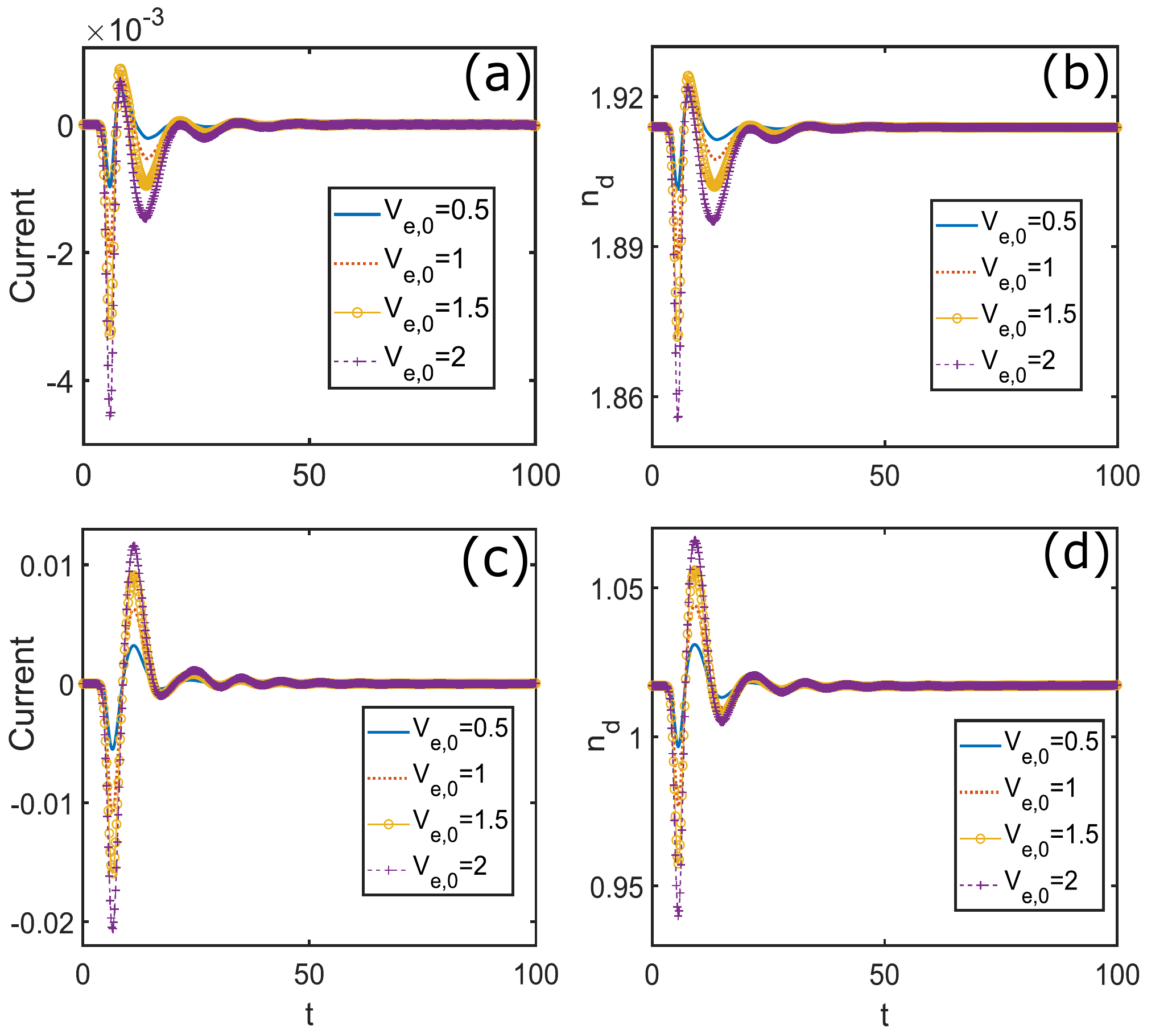}
\caption{The transient current and the occupation number in the double
occupation and Kondo regimes, where $\protect\varepsilon_d=-0.5$, $\Gamma
=0.04$, $t_{c}=5$, $\protect\alpha =\protect\omega _{d}=1$, and $t_{0}$ is
taken as the unit. (a)-(b) $U=0.05$; (c)-(d) $U=1$.}
\label{tdIwithoutphonon}
\end{figure}

\begin{figure}[tbp]
\includegraphics[width=0.9\linewidth]{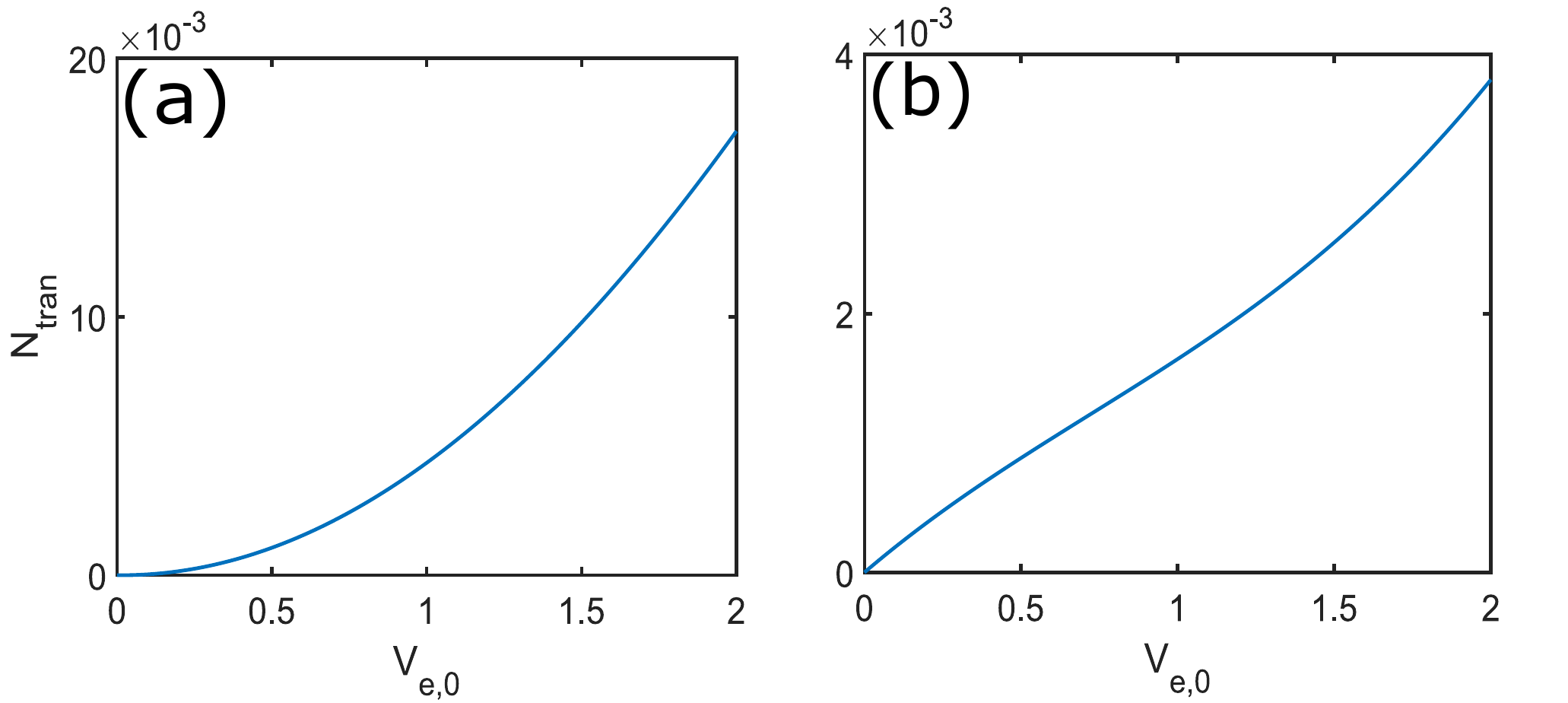}
\caption{The net transported charge in the double occupation and Kondo
regimes, where each lead contains 100 sites, $\protect\varepsilon_d=-0.5$, $%
\Gamma =0.04$, $t_{c}=5$, $\protect\alpha =\protect\omega _{d}=1$, and $%
t_{0} $ is taken as the unit. (a) $U=0.05$; (b) $U=1$. }
\label{Ntrwithoutphonon}
\end{figure}

For the system with bias time (in)dependent $V_{\mathrm{e}}(t)$, we can
study the occupation number $\left\langle n_{d}\right\rangle $, the
displacement $\left\langle R\right\rangle $, and the correlation function $%
C_{\alpha }$. In the non-equilibrium state, the current is defined as $%
I=\sum_{j,\sigma }\partial _{t}\left\langle c_{j,\sigma ,\mathrm{L}%
}^{\dagger }c_{j,\sigma ,\mathrm{L}}\right\rangle $. The Heisenberg equation
of motion leads to%
\begin{eqnarray}
I &=&-V\text{Im}\{e_{\lambda }[\left\langle P_{z}(f_{\gamma }^{\dagger
}+f_{\gamma })(\gamma c_{0,\uparrow ,\mathrm{L}}-c_{0,\downarrow ,\mathrm{L}%
})\right\rangle _{\mathrm{GS}} \\
&&+\left\langle (f_{\gamma }^{\dagger }+f_{\gamma })c_{0,\uparrow ,\mathrm{L}%
}\right\rangle _{\mathrm{GS}}-\gamma \left\langle (f_{\gamma }^{\dagger
}-f_{\gamma })c_{0,\downarrow ,\mathrm{L}}\right\rangle _{\mathrm{GS}}]\},
\notag
\end{eqnarray}%
where the average values depend on the covariance matrix $\Gamma _{m}$ (see
Appendix A). The derivatives $\sigma _{c}=\partial _{V_{e}}I$ give the
conductance.

\subsection{Transport in the Anderson model}

In this subsection, we present results for electron transport in the
Anderson model without the electron-phonon interaction. We compute the
non-linear conductance in both the Kondo and doubly occupied regimes.

For the Kondo regime we set parameters $(U,\varepsilon _{d},\Gamma
)=(1,-0.5,0.16)$ and for the doubly occupied regime we choose $%
(U,\varepsilon _{d},\Gamma )=(0.05,-0.5,0.04)$, Fig. \ref{tdwithoutphonon}
shows the occupation number $n_{d}$ and the conductance $\sigma _{c}$ in the
steady state as the function of $V_{e}$. Fig. \ref{tdwithoutphonon}a shows
the conductance $\sigma _{c}$ around zero bias in the Kondo regime. The peak
value $\sigma _{c}=2$ originates from the Kondo resonance and agrees with
the Friedel sum rule. In the doubly occupied regime, when the bias $V_{%
\mathrm{e}}$ crosses $\sim \varepsilon _{d}+U$, the energy level of the
doubly occupied state in HOMO is higher than the Fermi surface of the right
reservoir. As a result, the transport channel is turned on and the electron
in HOMO tunnels to the reservoir, resulting in the decay of $n_{d}$ and the
appearance of a conductance peak around $\varepsilon _{d}+U$ in Fig. \ref%
{tdwithoutphonon}b.

In the experiment, an ultrafast pulse is applied to shift the chemical
potential of the right lead, where the HOMO is doubly occupied. The effect
of the ultrafast pulse centered at the instant $t_{c}$ is described by the
time-dependent bias%
\begin{equation}
V_{\mathrm{e}}(t)=-V_{\mathrm{e},0}e^{-\alpha ^{2}(t-t_{c})^{2}}\sin \omega
_{d}(t-t_{c}),  \label{Ve}
\end{equation}%
where $V_{\mathrm{e},0}$ is the intensity, $\alpha $ determines the width of
the pulse, and $\omega _{d}$ is the frequency.

For different pulse intensities $V_{\mathrm{e},0}$, the transient current $I$
and the occupation number $n_{d}$ as a function of time $t$ are shown in the
left and right panels of Fig. \ref{tdIwithoutphonon}. The two rows in Fig. %
\ref{tdIwithoutphonon} correspond to system parameters $(U,\varepsilon
_{d})=(0.05,-0.5)$ and $(1,-0.5)$ in the double occupation and Kondo
regimes, where $\Gamma =0.04$, $t_{c}=5$, $\alpha =\omega _{d}=1$, and $%
t_{0} $ is taken as the unit. The number $N_{\mathrm{tran}}$ of electrons
transferred from the left lead to the right lead as the function of pulse
intensity $V_{e,0}$ is displayed in Fig. \ref{Ntrwithoutphonon}.

\begin{figure*}[tbp]
\begin{center}
\includegraphics[width=0.9\linewidth]{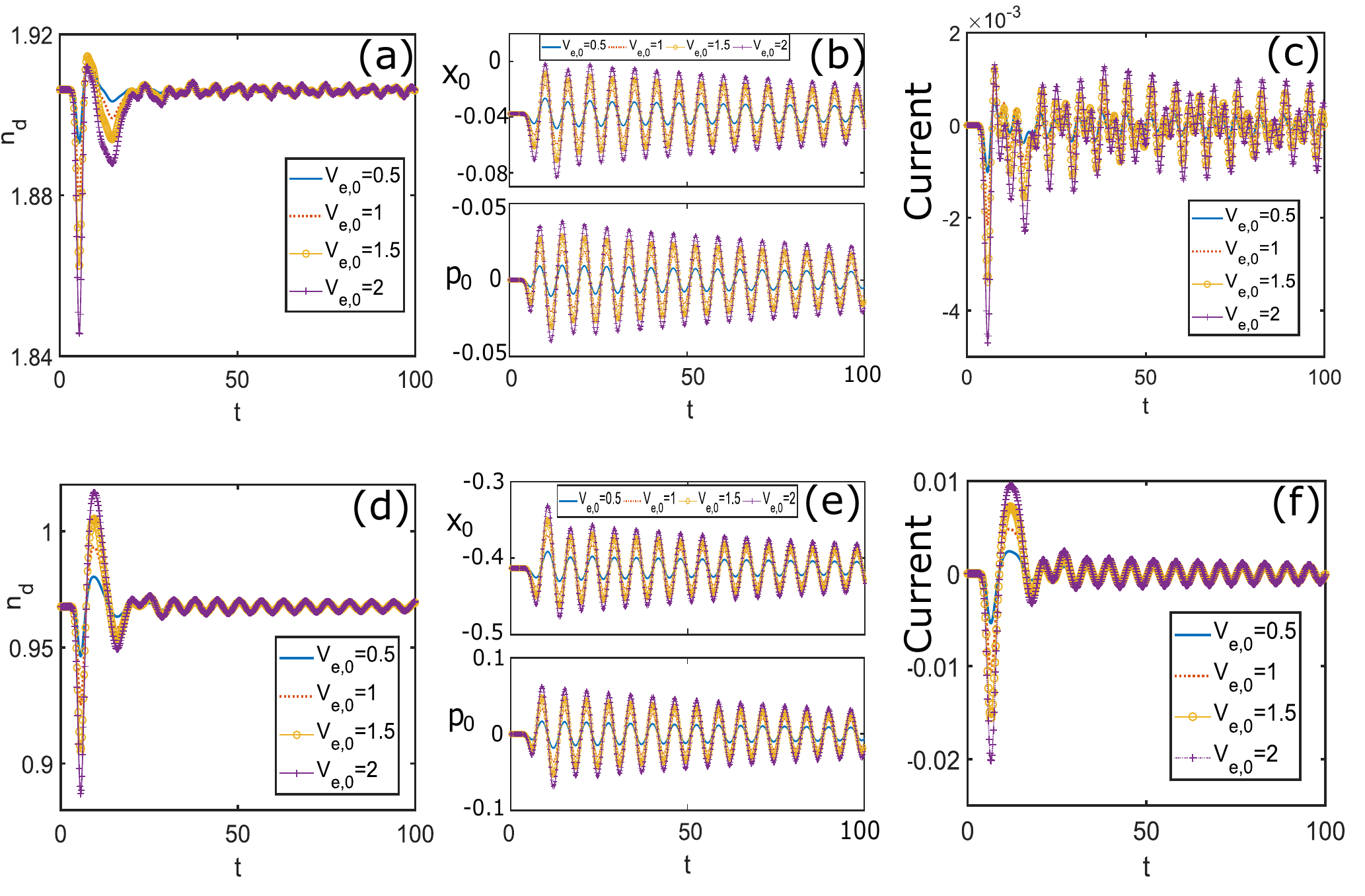}
\end{center}
\caption{The occupation number (the first column), the average value of the
quadrature (the second column), and the transient current (the third column)
in the double occupation (the first row) and the Kondo (the second row)
regimes, where $\protect\varepsilon_d=-0.5$, $\Gamma =0.04$, $g=0.2$, $%
t_{c}=5$, $\protect\alpha =\protect\omega _{d}=1$, and $t_{0}$ is taken as
the unit. (a)-(c): $U=0.05$; (d)-(f): $U=1$.}
\label{D1withphonon}
\end{figure*}

The pulse has a single period Sine-type shape, which first shifts the Fermi
level of the right reservoir downwardly and then lifts it above the Fermi
level of the left reservoir after $t_{c}$. In the double occupation regimes,
as shown by the first row of Fig. \ref{tdIwithoutphonon}, when the Fermi
energy is resonant with $\varepsilon _{d}+U<0$ at the instant $t_{\mathrm{res%
}}<t_{c}$, the transport channel is turned on and the transient current
flowing to the right reservoir establishes. However, after the instant $%
t_{c} $ when the Fermi level of the right reservoir is higher than that of
the left reservoir, there is no resonant energy level. Eventually, due to
the asymmetric spectral structure around the Fermi level $\varepsilon _{F}=0$%
, the light pulse induces the non-zero net charge transported $\sim 10^{-2}$
from the left to the right, as shown in Fig. \ref{Ntrwithoutphonon}a. In the
Kondo regime, the energy spectrum is symmetric around the Fermi level, as a
result, the net charge transported is highly reduced. We note that a finite
value of the net charge $\sim 10^{-3}$ in Fig. \ref{Ntrwithoutphonon}b most
likely arises from the nonlinear electronic dispersion relation in
reservoirs.

\subsection{Transport in the Anderson-Holstein model}

\begin{figure}[tbp]
\includegraphics[width=0.9\linewidth]{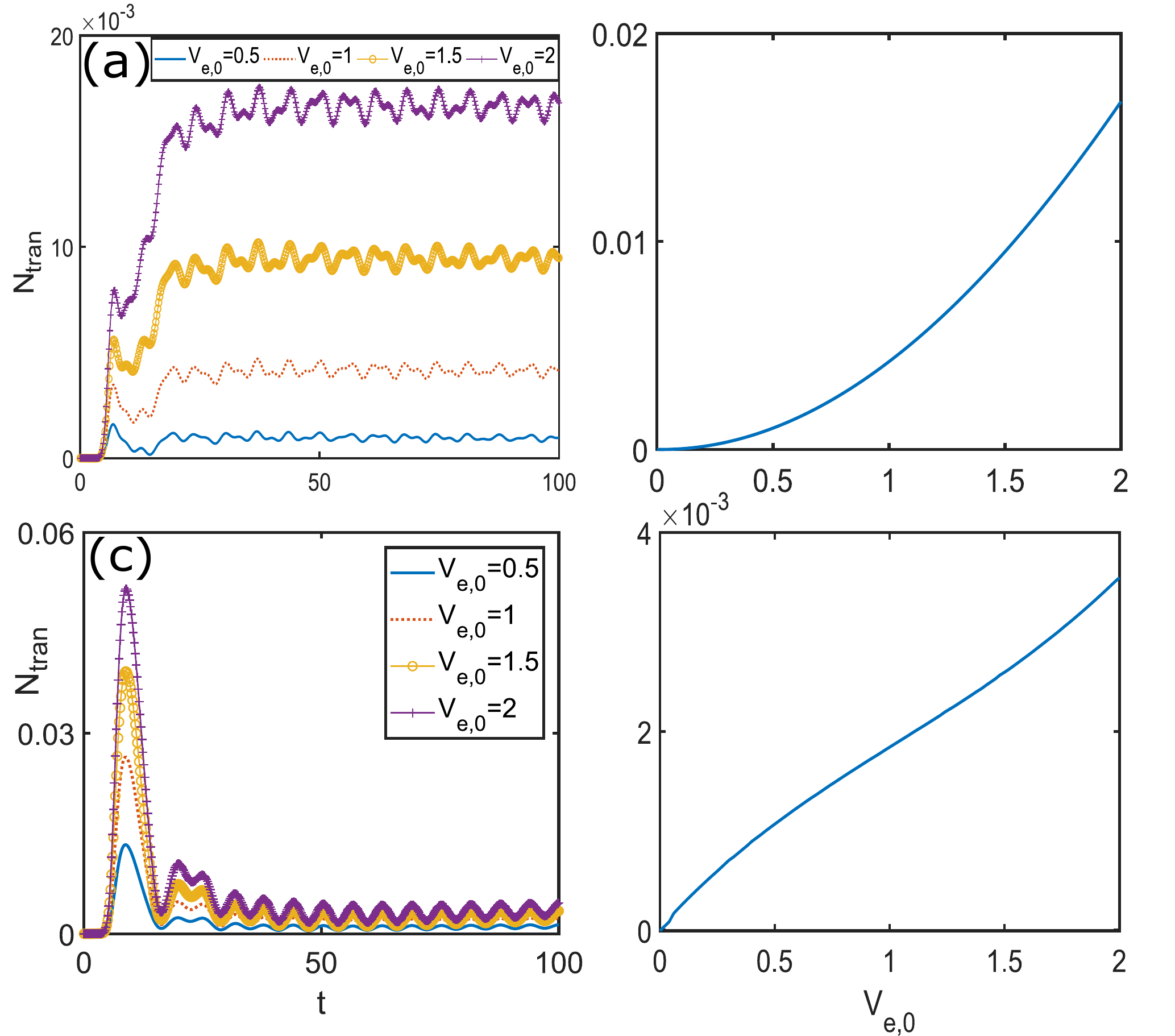}
\caption{The net charge transport in the double occupation (the first row)
and Kondo regimes (the second row), where $\protect\varepsilon_d=-0.5$, $%
\Gamma =0.04$, $g=0.2$, $t_{c}=5$, $\protect\alpha =\protect\omega _{d}=1$,
and $t_{0}$ is taken as the unit. (a)-(b): The time-dependent and average
value of net charges transport to the right reservoir in the double
occupation regime $U=0.05$; (c)-(d): The time-dependent and average value of
net charges transport to the right reservoir in the Kondo regime $U=1$.}
\label{Ntrwithphonon}
\end{figure}

In this subsection, we concentrate on the phonon excitations generated by
the ultrafast pulse in the doubly occupied regime. Since the light pulse
induces a transient current, i.e., the transport of electrons between HOMO
and leads, the holes are created in the HOMO. The Coulomb interaction
between the charge in the substrate and the hole excitation in HOMO induces
the molecular vibration after interacting with the ultrafast pulse, which is
described by the time-dependent average value $\left\langle R\right\rangle
=(x_{0},p_{0})^{T}$ of the quadrature.

In the first and second rows of Fig. \ref{D1withphonon}, we show the
occupation number $n_{d}$, the average value $\left\langle R\right\rangle
=(x_{0},p_{0})^{T}$ of the quadrature, and the transient current $I$ in the
double occupation and Kondo regimes $(U,\varepsilon _{d})=(0.05,-0.5)$ and $%
(1,-0.5)$, respectively. Here, $\Gamma =0.04$, $g=0.2$, $t_{c}=5$, $\alpha
=\omega _{d}=1$, and $t_{0}$ is taken as the unit. The electrons are
transferred from the left reservoir to the right one when the terahertz
pulse is applied, and the long-lived oscillation of net charge transport
around a center value is observed, as shown in the left panel of Fig. \ref%
{Ntrwithphonon} for the double occupation and Kondo regimes. Here, the
center value of the oscillation as a function of pulse intensity is also
shown in the right panel of Fig. \ref{Ntrwithphonon}.

In the initial stage, i.e., $t<t_{c}$, the ultrafast pulse shifts the Fermi
level of the right reservoir downwardly, thus electrons in HOMO flow to the
right reservoir, and the occupation number $n_{d}$ decreases, as shown in
the first column of Fig. \ref{D1withphonon}. Since more holes are generated
in the HOMO, the molecule is driven away from the equilibrium position along
the negative direction by the electron-phonon interaction, as shown in the
second column of Fig. \ref{D1withphonon}. When the Fermi level of the right
reservoir is resonant with $\tilde{\varepsilon}_{d}$, the transport channel
is turned on, and the significant transient current is formed, as shown in
the third column of Fig. \ref{D1withphonon} and the left panel of Fig. \ref%
{Ntrwithphonon}.

In the intermediate stage, i.e., $t_{c}<t<t_{c}+\pi /\omega _{d}$, since the
Fermi level of the right reservoir is above that of the left one, the HOMO
is repumped, as shown in the first column of Fig. \ref{D1withphonon}.
However, the resonant transport is absent since the Fermi level of the right
reservoir is detuned from LUMO. Eventually, less electrons flow back to the
left reservoir, and the finite net charge is transported to the right
reservoir, as shown in the Fig. \ref{Ntrwithphonon}. In the final stage $%
t>t_{c}+\pi /\omega _{d}$, when the light pulse is turned off, the
long-lived phonon mode with frequency $\sim \omega _{b}$ survives, as shown
in Fig. \ref{D1withphonon}b. Due to the indirect coupling to the reservoir
through electrons in HOMO, the phonon mode has a finite life-time, which is
revealed by the slowly decaying amplitude of the oscillation in \ref%
{D1withphonon}b. The molecular vibration induces the long-lived oscillation
of current and net charge transport around their values in the steady state,
as shown in the third column of Fig. \ref{D1withphonon} and the left panel
of Fig. \ref{Ntrwithphonon}.

\subsection{Discussion of results}

Before concluding this section we would like to highlight several results of
our analysis.

The main difference between the regimes of single and double occupancy is
the dependence of the net transferred charge on the amplitude of the THz
light pulse $V_{e,0}$. In the former case we observe a linear dependence of
the transferred charge on $V_{e,0}$ (see Fig. 8b), whereas in the latter
case we find a non-linear dependence (see Fig. 8a). A special feature of the
single occupancy regime is the existence of the Kondo resonance at the Fermi
energy, which gives rise to finite DC conductance. This however is not the
whole story. In Appendix E we present the analysis of the photoinduced
current in the RLM with the energy of the localized state set exactly at the
Fermi energy, $\epsilon _{d}=0$, which would also allow for finite DC
conductance. In the RLM we also find a non-linear dependence of the
photocurrent on $V_{e,0}$ (see fig. 12d). We remind the readers that that
the integral of the time dependent electric field in the THz pulse is zero.
Hence for a completely linear system the net transferred charge should
vanish. The linear dependence of photocurrent on $V_{e,0}$ in the Kondo
regime is thus a surprising feature of the system.

Another interesting result of our analysis is the existence of different
time scales characterizing the transient dynamics. The occupation number of
electrons on the localized orbital exhibits two distinct scales in its
dynamics. There is a relatively short timescale over which the strong
amplitude deviation from equilibrium configuration decays. It is followed by
small amplitude oscillations that match the frequency of the photo-excited
phonon mode, with both oscillations of the phonon amplitude and the electron
occupation number exhibiting slow decay. The existence of different time
scales is a common feature of pump and probe experiments in strongly
correlated electron systems (see e.g. Ref. \cite{Cao2018}). Slow relaxation
of phonon excitations was one of the key features observed in experiments by
Cocker \textit{et al.} \cite{Cocker2016}. Our analysis suggests that the
origin of this relaxation is due to phonon displacements modifying virtual
tunneling processes of reservoir electrons into the localized orbital, which
ultimately allows the phonon energy to be converted into particle-hole
excitations.

\section{Summary and Outlook}

We introduced a new method for analyzing the Anderson-Holstein impurity
model. Key ingredients of the approach are two unitary transformations that
use the parity conservation to partially decouple the impurity degrees of
freedom and a generalized polaron transformation to entangle phonons and
electrons. An appealing aspect of the new method is that degrees of freedom
at very different energy scales, from the local repulsion $U$ to the Kondo
temperature $T_{K}$, are described within the same framework and without the
numerical demands of the NRG and DMRG calcualtions. To verify the accuracy
of the new approach we computed the properties of the Anderson model,
including the equilibrium spectral function, linear and non-linear DC
conductance. We demonstrated that we correctly reproduce known results for
this model. We extended the analysis of the Anderson model to include
electron-phonon interactions and showed that it leads to a suppression of
the Kondo resonance. We used our approach to analyze THz-STM experiments of
tunneling through a single molecule. We found that a picosecond light pulse
that induces the current flow gives rise to strong oscillations of the
molecular phonon, which persist long after the end of the pulse. We analyzed
the time dependence of the current induced by the THz pulse and showed a
strong difference between the Kondo and doubly occupied regimes of the
molecule.

Our work can be extended in several directions. One interesting question is
developing a deeper understanding of the interplay of electron-electron and
electron-phonon interactions in pump and probe experiments \cite%
{Yang2007,Stojchevska2010,Pomarico2017,Schmitt2008,Schmitt2011}. Recent
time-resolved ARPES/Xray experiments by Gerber et al. measured changes of
the electron energy relative to the phonon displacement in a photoexcited
iron selenide \cite{Gerber2017}. They observed a strong renormalization of
the electron-phonon coupling relative to the value predicted from band
structure calculations and attributed it to the effects of electron-electron
interactions. Extension of the formalism presented in this paper can be used
to study the time dependent spectral function of electrons following a THz
light pulse. This analysis will provide a direct comparison between changes
of the electron energy and the phonon amplitude.

Another interesting question is the analysis of shot to shot fluctuations in
the number of electrons transferred through the junction during the THz
pulse. In the case of DC transport the study of shot noise has been a
valuable tool for understanding underlying many-body states (see Refs. \cite%
{Nazarov,Blanter2000,Beenakker2003} for a review), including the
demonstration of fractional statistics in the fractional quantum Hall states
\cite{Kane1993,Reznikov2002} and understanding the back-scattering mechanism
in the Kondo system \cite{Sela2006,Golub2006,Yamauchi2011} as a probe of the
low temperature fixed point.

The formalism developed in our paper introduces a powerful new technique
that can be used for the theoretical analysis of many different types of
nonequilibrium problems. For example, it should provide useful insights into
the analysis of resonant XRay scattering experiments \cite{Ament2011} in
materials with mobile electrons. The essence of these experiments is that an
incident photon excites an electron from the core orbital into one of the
unoccupied bands, which results in a sudden introduction of a positively
charged hole in the sea of conduction band electrons\cite{Benjamin2013}. The
attractive potential of the hole is strong enough to allow the formation of
the excitonic bound state, however, such bound state can be occupied by only
one of the electrons. Hence, an analysis of the many-body dynamics following
absorption of the incident photon should include the local repulsion on the
hole site. The formalism developed in this paper can be used for computing
RIXS processes involving the excitation of bosonic modes, such as phonons
and magnons, as well as a continuum of electron-hole pairs. Another
interesting direction for applying the ideas presented in this paper is to
use variational non-Gaussian states as a solver for non-equilibrium DMFT
calculations \cite{Aoki2014,Ganahl2015}.

\section{Acknowledgements}

We are grateful for useful discussions with D. Abanin, Y. Ashida, T. Cocker,
A. Georges, L. Glazman, M. Kanasz-Nagy, A. Lichtenstein, A. Millis, A.
Rosch, A. Rubtsov, Z.X. Shen, J. van den Brink, G. Zarand. T. S. acknowledges the
Thousand-Youth-Talent Program of China. J.I.C acknowledges the ERC Advanced
Grant QENOCOBA under the EU Horizon2020 program (grant agreement 742102) and
the German Research Foundation (DFG) under Germany's Excellence Strategy
through Project No. EXC-2111-390814868 (MCQST) and within the D-A-CH
Lead-Agency Agreement through project No. 414325145 (BEYOND C). ED
acknowledges support from the Harvard-MIT CUA, Harvard-MPQ Center,
AFOSR-MURI: Photonic Quantum Matter (award FA95501610323), and DARPA DRINQS
program (award D18AC00014).

{\tiny \
\bibliographystyle{unsrt}
\bibliography{refs_Anderson_paper_intro,refs_Anderson_paper_summary}
}

\begin{widetext}
\appendix

\section{Mean-field Hamiltonian of fermions}

The Wick-theorem gives the mean-field Hamiltonian%
\begin{equation}
\mathcal{H}_{m}=-i\frac{1}{2}W_{f}\left(
\begin{array}{cc}
\mathcal{E}_{0} & \Delta _{0} \\
\Delta _{0}^{\dagger } & -\mathcal{E}_{0}^{T}%
\end{array}%
\right) W_{f}^{\dagger }+\mathcal{H}_{m}^{P},  \label{Hm}
\end{equation}%
in the Majorana basis, where the matrices%
\begin{eqnarray}
\mathcal{E}_{0} &=&\frac{1}{2}\left(
\begin{array}{ccc}
\tilde{U} & v_{-} & v_{-} \\
v_{-}^{\dagger } & 2h_{\mathrm{L}} & 0 \\
v_{-}^{\dagger } & 0 & 2h_{\mathrm{R}}%
\end{array}%
\right) ,\Delta _{0}=\frac{1}{2}\left(
\begin{array}{ccc}
0 & -v_{+} & -v_{+} \\
v_{+}^{T} & 0 & 0 \\
v_{+}^{T} & 0 & 0%
\end{array}%
\right) ,  \notag \\
\mathcal{E}_{P} &=&\frac{1}{2}\gamma \left(
\begin{array}{ccc}
U_{P} & v_{-} & v_{-} \\
-v_{+}^{T} & 0 & 0 \\
-v_{+}^{T} & 0 & 0%
\end{array}%
\right) ,\Delta _{P}=\frac{1}{2}\gamma \left(
\begin{array}{ccc}
0 & v_{-} & v_{-} \\
0 & 0 & 0 \\
0 & 0 & 0%
\end{array}%
\right)
\end{eqnarray}%
are defined by $U_{P}=\tilde{U}+2\tilde{\varepsilon}_{d}-2\Delta
_{R}^{T}G_{\lambda }$, the vectors $v_{\pm }=V\left\langle e^{\mp
iR^{T}\lambda }\right\rangle _{\mathrm{GS}}(1,\pm \gamma )\otimes
(1,0_{N-1}) $, and the hopping matrices $h_{\mathrm{a}=\mathrm{L,R}%
}=I_{2}\otimes (-t_{0}\delta _{i,j\pm 1}-\mu _{\mathrm{a}}\delta _{ij})$ in
the left and right reservoirs. The matrix $\mathcal{H}_{m}^{P}=4\delta
E_{P}/\delta \Gamma _{m}$ is determined by the derivative of $%
E_{P}=\left\langle P_{z}c^{\dagger }\mathcal{E}_{P}c\right\rangle +2$Re$%
\left\langle P_{z}c^{T}\Delta _{P}c\right\rangle $.

The $2\times 2$ Gram matrix%
\begin{eqnarray}
\mathbf{G} &=&\left\langle P_{c}P_{c}\right\rangle \Gamma _{b}+\left\langle
P_{h}P_{h}\right\rangle (4\sigma ^{y}\lambda \lambda ^{T}\sigma ^{y}  \notag
\\
&&+\Delta _{R}\Delta _{R}^{T}+2i\Delta _{R}\lambda ^{T}\sigma ^{y}-2i\sigma
^{y}\lambda \Delta _{R}^{T})
\end{eqnarray}%
of the tangential vector $U_{\mathrm{ph}}U_{\mathrm{A}}U_{\mathrm{GS}%
}V_{h}\left\vert 0\right\rangle $ is determined by%
\begin{eqnarray}
\left\langle P_{c}P_{c}\right\rangle &=&\left\langle f_{\gamma }^{\dagger
}f_{\gamma }\right\rangle _{\mathrm{GS}}-\left\langle P_{z}f_{\gamma
}^{\dagger }f_{\gamma }\right\rangle _{\mathrm{GS}}^{2},  \notag \\
\left\langle P_{h}P_{h}\right\rangle &=&\left\langle P_{c}P_{c}\right\rangle
-\frac{1}{2}\left\langle P_{z}f_{\gamma }^{\dagger }f_{\gamma }\text{:}%
C^{\dagger }O_{P}C\text{:}\right\rangle _{\mathrm{GS}}.
\end{eqnarray}%
The vector%
\begin{eqnarray}
\xi _{\tau } &=&i(\Gamma _{b}-\sigma ^{y})(\left\langle
P_{c}P_{c}\right\rangle G_{\lambda }-\gamma \Sigma _{c}\lambda )  \notag \\
&&-(2\sigma ^{y}\lambda +i\Delta _{R})(\frac{1}{2}U_{P}\left\langle
f_{\gamma }^{\dagger }f_{\gamma }\right\rangle _{\mathrm{GS}}+\gamma \Sigma
_{h})
\end{eqnarray}
is the overlap of the tangential vector and the right hide side of Eq. (\ref%
{IM}), where%
\begin{eqnarray}
\Sigma _{c} &=&i\frac{V}{2}\sum_{\mathrm{a}}[e_{\lambda }\left\langle
(1+\gamma P_{z})f_{\gamma }^{\dagger }(\gamma c_{0,\uparrow ,\mathrm{a}%
}-c_{0,\downarrow ,\mathrm{a}})\right\rangle _{\mathrm{GS}}  \notag \\
&&+e_{\lambda }^{\ast }\left\langle (1-\gamma P_{z})(\gamma c_{0,\uparrow ,%
\mathrm{a}}^{\dagger }+c_{0,\downarrow ,\mathrm{a}}^{\dagger })f_{\gamma
}^{\dagger }\right\rangle _{\mathrm{GS}}] \\
&&+\left\langle P_{z}f_{\gamma }^{\dagger }f_{\gamma }\right\rangle _{%
\mathrm{GS}}\text{Im}\left\langle \mathcal{I}_{dc}\right\rangle _{\mathrm{GS}%
},  \notag
\end{eqnarray}%
\begin{eqnarray}
\Sigma _{h} &=&\frac{1}{2}V\sum_{\mathrm{a}}[e_{\lambda }\left\langle
f_{\gamma }^{\dagger }(\gamma c_{0,\uparrow ,\mathrm{a}}-c_{0,\downarrow ,%
\mathrm{a}})\right\rangle _{\mathrm{GS}}  \notag \\
&&-e_{\lambda }^{\ast }\left\langle (\gamma c_{0,\uparrow ,\mathrm{a}%
}^{\dagger }+c_{0,\downarrow ,\mathrm{a}}^{\dagger })f_{\gamma }^{\dagger
}\right\rangle _{\mathrm{GS}}] \\
&&-E_{P}\left\langle P_{z}f_{\gamma }^{\dagger }f_{\gamma }\right\rangle _{%
\mathrm{GS}}-\frac{1}{2}\left\langle P_{z}f_{\gamma }^{\dagger }f_{\gamma }%
\text{:}C^{\dagger }\mathcal{H}_{f}^{P}C\text{:}\right\rangle _{\mathrm{GS}},
\notag
\end{eqnarray}%
and $\mathcal{H}_{f}^{P}=iW_{f}^{\dagger }\mathcal{H}_{m}^{P}W_{f}/2$ is the
matrix in the Nambu basis.

The mean-field Hamiltonian, the Gram matrix, and the vector $\xi _{\tau }$
involve the average values $\left\langle P_{z}\right\rangle _{\mathrm{GS}}$,
$\left\langle P_{z}C_{i}C_{j}^{\dagger }\right\rangle _{\mathrm{GS}}$ and $%
\left\langle P_{z}f_{\gamma }^{\dagger }f_{\gamma }C_{i}^{\dagger
}C_{j}\right\rangle _{\mathrm{GS}}$ on the Gaussian state. We consider the
average values of operators $P_{z}^{\theta }=e^{\theta f^{\dagger }f}P_{z}$
and $P_{z}^{\theta }C_{i}C_{j}^{\dagger }$, which eventually give $%
\left\langle P_{z}\right\rangle _{\mathrm{GS}}=\left\langle P_{z}^{\theta
=0}\right\rangle $ and $\left\langle P_{z}C_{i}C_{j}^{\dagger }\right\rangle
_{\mathrm{GS}}=\left\langle P_{z}^{\theta =0}C_{i}C_{j}^{\dagger
}\right\rangle _{\mathrm{GS}}$, as well as $\left\langle P_{z}f_{\gamma
}^{\dagger }f_{\gamma }C_{i}^{\dagger }C_{j}\right\rangle _{\mathrm{GS}}$ by
the derivative $\left. \partial _{\theta }\left\langle P_{z}^{\theta
}C_{i}C_{j}^{\dagger }\right\rangle _{\mathrm{GS}}\right\vert _{\theta =0}$.

As shown in Ref. \cite{Shi2018,Ashida2018}, the average values $\left\langle P_{z}^{\theta
}\right\rangle _{\mathrm{GS}}=-$Pf$(\Gamma _{F}/2)$ and%
\begin{eqnarray}
\left\langle P_{z}^{\theta }c_{l}^{\dagger }c_{k}\right\rangle _{\mathrm{GS}%
} &=&\frac{1}{4}\left\langle P_{z}^{\theta }\right\rangle _{\mathrm{GS}%
}[\left( 1,i\right) \mathcal{S}\Theta \left(
\begin{array}{c}
1 \\
-i%
\end{array}%
\right) ]_{kl},  \notag \\
\left\langle P_{z}^{\theta }c_{l}c_{k}\right\rangle _{\mathrm{GS}} &=&-\frac{%
1}{4}\left\langle P_{z}^{\theta }\right\rangle _{\mathrm{GS}}[(1,i)\mathcal{S%
}\left(
\begin{array}{c}
1 \\
i%
\end{array}%
\right) ]_{kl},
\end{eqnarray}%
are determined by the Pfaffian of%
\begin{equation}
\Gamma _{F}=\sqrt{1+\Theta }\Gamma _{m}\sqrt{1+\Theta }-\sigma (1-\Theta ),
\end{equation}%
$\mathcal{S}=(\sigma \Gamma _{m}-1)\mathcal{T}$, and%
\begin{equation}
\mathcal{T}=\frac{1}{1+\frac{1}{2}(1+\Theta )(\sigma \Gamma _{m}-1)},
\end{equation}%
where $\Theta =I_{2}\otimes \sigma _{\theta }$, $\sigma _{\theta
}=diag(-e^{\theta },1_{1\times N},-1_{1\times N},1_{1\times N},-1_{1\times
N})$ is a diagonal matrix and%
\begin{equation}
\sigma =\left(
\begin{array}{cc}
{0} & {\openone}_{4N+1} \\
-{\openone}_{4N+1} & 0%
\end{array}%
\right)
\end{equation}%
is a symplectic matrix. The anti-commutation relation results in%
\begin{eqnarray}
\left\langle P_{z}^{\theta =0}c_{l}c_{k}^{\dagger }\right\rangle _{\mathrm{GS%
}} &=&\left\langle P_{z}^{0}\right\rangle \delta _{lk}-\left\langle
P_{z}^{0}c_{k}^{\dagger }c_{l}\right\rangle ,  \notag \\
\left\langle P_{z}^{\theta =0}c_{l}^{\dagger }c_{k}^{\dagger }\right\rangle
_{\mathrm{GS}} &=&(\sigma _{0})_{kk}[\left\langle
P_{z}^{0}c_{k}c_{l}\right\rangle ]^{\ast }(\sigma _{0})_{ll}.
\end{eqnarray}

By the derivative to $\theta $ and taking the limit $\theta \rightarrow 0$,
we obtain%
\begin{equation}
\left\langle P_{z}f_{\gamma }^{\dagger }f_{\gamma }c_{i}^{\dagger
}c_{j}\right\rangle =-\frac{1}{2}g_{F}\left\langle P_{z}c_{i}^{\dagger
}c_{j}\right\rangle _{\mathrm{GS}}-\frac{1}{4}\left\langle
P_{z}\right\rangle _{\mathrm{GS}}[\left( 1,i\right) \mathcal{S}w(1-\frac{1}{2%
}\mathcal{S}\Theta )\left(
\begin{array}{c}
1 \\
-i%
\end{array}%
\right) ]_{ji},
\end{equation}%
and%
\begin{equation}
\left\langle P_{z}f_{\gamma }^{\dagger }f_{\gamma }c_{i}c_{j}\right\rangle =-%
\frac{1}{2}g_{F}\left\langle P_{z}c_{i}c_{j}\right\rangle _{\mathrm{GS}}-%
\frac{1}{8}\left\langle P_{z}\right\rangle _{\mathrm{GS}}[(1,i)\mathcal{S}w%
\mathcal{S}\left(
\begin{array}{c}
1 \\
i%
\end{array}%
\right) ]_{ji},
\end{equation}%
where $w=I_{2}\otimes diag(1,0_{1\times 4N})$ and%
\begin{equation}
g_{F}=tr(\Gamma _{F}^{-1}\sigma w+\frac{1}{2}\mathcal{T}\sigma w\Gamma _{m}).
\end{equation}%
The commutation relation leads to%
\begin{eqnarray}
\left\langle P_{z}f_{\gamma }^{\dagger }f_{\gamma }c_{i}c_{j}^{\dagger
}\right\rangle _{\mathrm{GS}} &=&\left\langle P_{z}f_{\gamma }^{\dagger
}f_{\gamma }\right\rangle _{\mathrm{GS}}\delta _{ij}-\left\langle
P_{z}f_{\gamma }^{\dagger }f_{\gamma }c_{j}^{\dagger }c_{i}\right\rangle _{%
\mathrm{GS}},  \notag \\
\left\langle P_{z}f_{\gamma }^{\dagger }f_{\gamma }c_{i}^{\dagger
}c_{j}^{\dagger }\right\rangle _{\mathrm{GS}} &=&(\sigma _{\theta
})_{ii}\left\langle P_{z}f_{\gamma }^{\dagger }f_{\gamma
}c_{j}c_{i}\right\rangle ^{\ast }(\sigma _{\theta })_{jj}  \notag \\
&&+\delta _{i1}\left\langle P_{z}f_{\gamma }^{\dagger }c_{j}^{\dagger
}\right\rangle -\left\langle P_{z}f_{\gamma }^{\dagger }c_{i}^{\dagger
}\right\rangle \delta _{j1}.
\end{eqnarray}

The mean-field Hamiltonian $\mathcal{H}_{m}^{P}$ is determined by the
derivatives%
\begin{equation}
\frac{\delta }{\delta \Gamma _{m,ij}}\left\langle P_{z}\right\rangle _{%
\mathrm{GS}}=-\frac{1}{2}\left\langle P_{z}\right\rangle _{\mathrm{GS}}\sqrt{%
1+\Theta }\frac{1}{\Gamma _{F}}\sqrt{1+\Theta },  \label{A1}
\end{equation}%
\begin{eqnarray}
\frac{\delta }{\delta \Gamma _{m,ij}}\left\langle P_{z}c_{l}^{\dagger
}c_{k}\right\rangle _{\mathrm{GS}} &=&-\frac{1}{2}\left\langle
P_{z}c_{l}^{\dagger }c_{k}\right\rangle _{\mathrm{GS}}(\sqrt{1+\Theta }\frac{%
1}{\Gamma _{F}}\sqrt{1+\Theta })_{ij}  \notag \\
&&-i\frac{1}{4}\left\langle P_{z}\right\rangle _{\mathrm{GS}}[\mathcal{T}%
\Theta \left(
\begin{array}{c}
1 \\
-i%
\end{array}%
\right) ]_{jl}[\left( 1,i\right) \mathcal{T}^{T}]_{ki},  \label{A2}
\end{eqnarray}%
and%
\begin{eqnarray}
\frac{\delta }{\delta \Gamma _{m,ij}}\left\langle
P_{z}c_{l}c_{k}\right\rangle &=&-\frac{1}{2}\left\langle
P_{z}c_{l}c_{k}\right\rangle _{\mathrm{GS}}(\sqrt{1+\Theta }\frac{1}{\Gamma
_{F}}\sqrt{1+\Theta })_{ij}  \notag \\
&&+i\frac{1}{4}\left\langle P_{z}\right\rangle _{\mathrm{GS}}[\mathcal{T}%
\left(
\begin{array}{c}
1 \\
i%
\end{array}%
\right) ]_{jl}[(1,i)\mathcal{T}^{T}]_{ki}.  \label{A3}
\end{eqnarray}

\section{Linearized ODE for $\partial _{\protect\tau }\protect\lambda $}

In this section, we reduce EOM (\ref{dL}), i.e.,%
\begin{equation}
\partial _{\tau }\lambda ^{T}\mathbf{M}\partial _{\tau }\lambda =\partial
_{\tau }\lambda ^{T}\xi _{\tau }.
\end{equation}%
to the linear ODE, where $\xi _{\tau }=(\xi _{\tau ,x},\xi _{\tau ,p})^{T}$
and%
\begin{equation}
\mathbf{M=}\left(
\begin{array}{cc}
\mathbf{M}_{11} & \mathbf{M}_{12} \\
\mathbf{M}_{21} & \mathbf{M}_{22}%
\end{array}%
\right) .
\end{equation}

Since the left hand side of Eq. (\ref{dL}) is always real, the imaginary
part of $\partial _{\tau }\lambda ^{T}\xi _{\tau }$ must vanish, which gives
rise to%
\begin{equation}
\partial _{\tau }\lambda _{x}\text{Im}\xi _{\tau ,x}+\partial _{\tau
}\lambda _{p}\text{Im}\xi _{\tau ,p}=0\text{.}
\end{equation}%
Four possibilities may happen: (a) Im$\xi _{\tau ,x}=$Im$\xi _{\tau ,p}=0$;
(b) Im$\xi _{\tau ,x}=0$, Im$\xi _{\tau ,p}\neq 0$; (c) Im$\xi _{\tau
,x}\neq 0$, Im$\xi _{\tau ,p}=0$; (d) Im$\xi _{\tau ,x}\neq 0$, Im$\xi
_{\tau ,p}\neq 0$.

For the case (a), EOM (\ref{dL}) is reduced to $\mathbf{M}\partial _{\tau
}\lambda =\xi _{\tau }$. For the case (b), EOM (\ref{dL}) becomes%
\begin{equation}
\partial _{\tau }\lambda _{p}=0,\mathbf{M}_{11}\partial _{\tau }\lambda
_{x}=\xi _{\tau ,x}.
\end{equation}%
For the case (c), EOM (\ref{dL}) is%
\begin{equation}
\partial _{\tau }\lambda _{x}=0,\mathbf{M}_{22}\partial _{\tau }\lambda
_{p}=\xi _{\tau ,p}.
\end{equation}%
For the last case, EOM (\ref{dL}) reads%
\begin{eqnarray}
\partial _{\tau }\lambda _{p} &=&-\frac{\text{Im}\xi _{\tau ,x}}{\text{Im}%
\xi _{\tau ,p}}\partial _{\tau }\lambda _{x},  \notag \\
\partial _{\tau }\lambda _{x} &=&\frac{1}{v_{\xi }^{T}\mathbf{M}v_{\xi }}%
v_{\xi }^{T}\xi _{\tau },
\end{eqnarray}%
where the vector $v_{\xi }=(1,-$Im$\xi _{\tau ,x}/$Im$\xi _{\tau ,p})^{T}$.

\section{Eveluation of average values in $G_{R}(t)$}

In this Appendix, we calculate the average values in the retarded Green
function $G_{R}(t)$, which are%
\begin{equation}
\left\langle e^{-iR^{T}(t)\lambda }e^{iR^{T}\lambda }\right\rangle _{\mathrm{%
GS}},\left\langle e^{iR^{T}\lambda }e^{-iR^{T}(t)\lambda }\right\rangle _{%
\mathrm{GS}}
\end{equation}%
for the phonon and%
\begin{equation}
\left\langle F(t)F^{\dagger }\right\rangle _{\mathrm{GS}},\left\langle
F^{\dagger }F(t)\right\rangle _{\mathrm{GS}}
\end{equation}%
for fermions.

We first consider the phonon part. The average value%
\begin{equation}
\left\langle e^{-iR^{T}(t)\lambda }e^{iR^{T}\lambda }\right\rangle _{\mathrm{%
GS}}=\text{ }_{b}\left\langle \bar{\Psi}_{\mathrm{GS}}\right\vert
e^{-i\omega _{\mathrm{re}}tb^{\dagger }b}\left\vert \bar{\Psi}_{\mathrm{GS}%
}\right\rangle _{b}
\end{equation}%
can be rewritten as the mean value of $e^{-i\omega _{\mathrm{re}}b^{\dagger
}bt}$ on the normalized Gaussian state%
\begin{equation}
\left\vert \bar{\Psi}_{\mathrm{GS}}\right\rangle
_{b}=e^{iR^{T}S_{p}^{T}\lambda }\left\vert 0\right\rangle _{b},
\end{equation}%
where the symplectic matrix $S_{p}$ diagonalizes the phonon mean-field
Hamiltonian $\Omega _{\mathrm{re}}$ as $S_{p}^{T}\Omega _{\mathrm{re}%
}S_{p}=\omega _{\mathrm{re}}I_{2}$.

The Gaussian state $\left\vert \bar{\Psi}_{\mathrm{GS}}\right\rangle _{b}$
is fully characterized by the average value $\Delta _{\lambda
}=-2S_{p}^{-1}i\sigma ^{y}\lambda $ of quadrature and the covariance matrix $%
\bar{\Gamma}_{b}=I$. The average value%
\begin{equation}
\left\langle e^{-iR^{T}(t)\lambda }e^{iR^{T}\lambda }\right\rangle _{\mathrm{%
GS}}=e^{-\frac{1}{4}\Delta _{\lambda }^{T}\Delta _{\lambda }(1-e^{-i\omega _{%
\mathrm{re}}t})}.
\end{equation}%
follows from the result in Ref. \cite{Shi2018}. Following the similar
procedure, one can obtain $\left\langle e^{iR^{T}\lambda
}e^{-iR^{T}(t)\lambda }\right\rangle _{\mathrm{GS}}$.

We analyze the fermionic part in the next step. The average value%
\begin{eqnarray}
\left\langle F(t)F^{\dagger }\right\rangle _{\mathrm{GS}} &=&\frac{1}{4}%
[u_{-}^{T}g_{1}(t)u_{-}+\gamma u_{-}^{T}g_{2}(t)u_{+}  \notag \\
&&+\gamma u_{+}^{T}g_{3}(t)u_{-}+g_{4}(t)]
\end{eqnarray}%
contains four terms $g_{1}(t)=\left\langle C(t)C^{\dagger }\right\rangle _{%
\mathrm{GS}}$, $g_{2}(t)=\left\langle C(t)P_{z}C^{\dagger }\right\rangle _{%
\mathrm{GS}}$, $g_{3}(t)=\left\langle P_{z}C(t)C^{\dagger }\right\rangle _{%
\mathrm{GS}}$, and
\begin{equation}
g_{4}(t)=\left\langle P_{z}(f+f^{\dagger })(t)P_{z}(f+f^{\dagger
})\right\rangle _{\mathrm{GS}},
\end{equation}%
where $u_{\pm }^{T}=(1,\pm 1)\otimes (1,0_{2N})$.

It follows from the Heisenberg equations of motion that the first three
terms obey%
\begin{eqnarray}
i\partial _{t}g_{1,2}(t) &=&\mathcal{H}_{f}g_{1,2}(t),  \notag \\
i\partial _{t}g_{3}(t) &=&g_{3}(t)\mathcal{H}_{f},
\end{eqnarray}%
where $\mathcal{H}_{f}=iW_{f}^{\dagger }\mathcal{H}_{m}W_{f}/2$. The
solutions%
\begin{eqnarray}
g_{1,2}(t) &=&e^{-i\mathcal{H}_{f}t}g_{1,2}(0),  \notag \\
g_{3}(t) &=&g_{3}(0)e^{-i\mathcal{H}_{f}t},
\end{eqnarray}%
are determined by the boundary values $g_{1}(0)=\left\langle CC^{\dagger
}\right\rangle _{\mathrm{GS}}$, $g_{2}=\left\langle CP_{z}C^{\dagger
}\right\rangle _{\mathrm{GS}}$, and $g_{3}(0)=\left\langle P_{z}CC^{\dagger
}\right\rangle _{\mathrm{GS}}$ are obtained analytically in Appendix A.

Applying the parity operator on the mean-field Hamiltonian, we can write the
fourth term%
\begin{equation}
g_{4}(t)=e^{iE_{\mathrm{MF}}^{e}t}\text{ }_{f}\left\langle \Psi _{\mathrm{GS}%
}\right\vert (f_{\gamma }+f_{\gamma }^{\dagger })e^{-i\frac{1}{2}C^{\dagger }%
\mathcal{H}_{\Theta }Ct}(f_{\gamma }+f_{\gamma }^{\dagger })\left\vert \Psi
_{\mathrm{GS}}\right\rangle _{f}
\end{equation}%
by $\mathcal{H}_{\Theta }=\Theta \mathcal{H}_{f}\Theta $, where $H_{\mathrm{%
MF}}^{e}\left\vert \Psi _{\mathrm{GS}}\right\rangle _{f}=E_{\mathrm{MF}%
}^{e}\left\vert \Psi _{\mathrm{GS}}\right\rangle _{f}$. Introducing the
unitary transformation $V_{\Theta }$: $V_{\Theta }^{\dagger }CV_{\Theta
}=U_{\Theta }C$, we obtain%
\begin{equation}
g_{4}(t)=e^{i\bar{\epsilon}t}\text{ }_{f}\left\langle \bar{\Psi}_{\mathrm{GS}%
}\right\vert e^{-ic^{\dagger }\epsilon ct}\left\vert \bar{\Psi}_{\mathrm{GS}%
}\right\rangle _{f}
\end{equation}%
where $\bar{\epsilon}=E_{\mathrm{MF}}^{e}+tr\epsilon /2$, $U_{\Theta }$
diagonalizes the matrix $\mathcal{H}_{\Theta }$ as%
\begin{equation}
U_{\Theta }^{\dagger }\mathcal{H}_{\Theta }U_{\Theta }=D=\left(
\begin{array}{cc}
\epsilon & 0 \\
0 & -\epsilon%
\end{array}%
\right) ,\epsilon _{j}\geq 0,
\end{equation}%
and $\left\vert \bar{\Psi}_{\mathrm{GS}}\right\rangle _{f}=V_{\Theta
}^{\dagger }(f_{\gamma }+f_{\gamma }^{\dagger })\left\vert \Psi _{\mathrm{GS}%
}\right\rangle _{f}$ is a normalized Gaussian state.

The Gaussian state $\left\vert \bar{\Psi}_{\mathrm{GS}}\right\rangle _{f}$
is fully characterized by the covariance matrix%
\begin{eqnarray}
\bar{\Gamma}_{f} &=&\text{ }_{f}\left\langle \bar{\Psi}_{\mathrm{GS}%
}\right\vert CC^{\dagger }\left\vert \bar{\Psi}_{\mathrm{GS}}\right\rangle
_{f}  \notag \\
&=&U_{\Theta }^{\dagger }[(\sigma _{x}\Gamma _{f}u_{+}u_{+}^{T}\Gamma
_{f}\sigma _{x})^{T}-\Gamma _{f}u_{+}u_{+}^{T}\Gamma _{f}+\Gamma
_{f}]U_{\Theta }.
\end{eqnarray}%
It follows from the result in Ref. that%
\begin{equation}
g_{4}(t)=-e^{i\bar{\epsilon}t}\text{Pf}(\frac{\bar{\Gamma}_{F}}{2}),
\end{equation}%
where%
\begin{equation}
\bar{\Gamma}_{F}=\sqrt{1-e^{-i\epsilon t}}\bar{\Gamma}_{m}\sqrt{%
1-e^{-i\epsilon t}}-\sigma (1+e^{-i\epsilon t})
\end{equation}%
is determined by $\bar{\Gamma}_{m}=i(W_{f}\bar{\Gamma}_{f}W_{f}^{\dagger
}-1) $. Following the same procedure, one can also obtain the average value $%
\left\langle F^{\dagger }F(t)\right\rangle _{\mathrm{GS}}$.

\section{Mean-field Hamiltonian of penalty term}

In this Appendix, we derive the mean-field Hamiltonian of the penalty term $%
H_{\Lambda }$ by the Wick theorem. In the explicit form, the Hamiltonian%
\begin{eqnarray}
H_{\Lambda }/\Lambda &=&(1-2\bar{N}_{\uparrow })\sum_{j,a}c_{j,\uparrow
,a}^{\dagger }c_{j,\uparrow ,a}+(\sum_{ia}c_{i,\uparrow ,a}^{\dagger
}c_{i,\uparrow ,a})^{2}  \notag \\
&&+(1-2\bar{N}_{\downarrow })\sum_{i,a}c_{i,\downarrow ,a}^{\dagger
}c_{i,\downarrow ,a}+(\sum_{i,a}c_{i,\downarrow ,a}^{\dagger
}c_{i,\downarrow ,a})^{2}  \notag \\
&&+(\bar{N}_{\downarrow }-\bar{N}_{\uparrow })\gamma P_{z}+\gamma
P_{z}c^{\dagger }\mathcal{E}_{P}^{\mu }c+2\gamma \sum_{j,a}P_{z}f_{\gamma
}^{\dagger }f_{\gamma }c_{j,\downarrow ,a}^{\dagger }c_{j,\downarrow ,a}
\notag \\
&&+\bar{N}_{\uparrow }^{2}+\bar{N}_{\downarrow }^{2}-\bar{N}_{\uparrow }-%
\bar{N}_{\downarrow }+1,
\end{eqnarray}%
where the matrix%
\begin{equation}
\mathcal{E}_{P}^{\mu }=\left(
\begin{array}{ccc}
1-2\bar{N}_{\downarrow } & 0 & 0 \\
0 & (\delta _{\sigma \uparrow }-\delta _{\sigma \downarrow })\delta _{\sigma
\sigma ^{\prime }}\delta _{ij} & 0 \\
0 & 0 & (\delta _{\sigma \uparrow }-\delta _{\sigma \downarrow })\delta
_{\sigma \sigma ^{\prime }}\delta _{ij}%
\end{array}%
\right) .
\end{equation}

The Wick theorem gives the mean-field Hamiltonian $H_{\Lambda ,\mathrm{MF}%
}=i $:$A^{T}\mathcal{H}_{\Lambda ,m}A$:$/4$, where%
\begin{equation}
\mathcal{H}_{\Lambda ,m}/\Lambda =-i\frac{1}{2}W_{f}\left(
\begin{array}{cc}
\mathcal{E}_{\mu } & \Delta _{\mu } \\
\Delta _{\mu }^{\dagger } & -\mathcal{E}_{\mu }%
\end{array}%
\right) W_{f}^{\dagger }+4\frac{\delta E_{P}^{\mu }}{\delta \Gamma _{m}}
\end{equation}%
is determined by%
\begin{eqnarray}
\mathcal{E}_{\mu } &=&2\delta _{ab}[(1-\bar{N}_{\uparrow
}+\sum_{lc}\left\langle c_{l,\uparrow ,c}^{\dagger }c_{l,\uparrow
,c}\right\rangle _{\mathrm{GS}})\delta _{\sigma \uparrow }  \notag \\
&&+(1-\bar{N}_{\downarrow }+\sum_{lc}\left\langle c_{l,\downarrow
,c}^{\dagger }c_{l,\downarrow ,c}\right\rangle _{\mathrm{GS}})\delta
_{\sigma \downarrow }]\delta _{\sigma \sigma ^{\prime }}\delta _{ij}  \notag
\\
&&-2(\left\langle c_{j,\uparrow ,b}^{\dagger }c_{i,\uparrow ,a}\right\rangle
_{\mathrm{GS}}\delta _{\sigma \uparrow }+\left\langle c_{j,\downarrow
,b}^{\dagger }c_{i,\downarrow ,a}\right\rangle _{\mathrm{GS}}\delta _{\sigma
\downarrow })\delta _{\sigma \sigma ^{\prime }},
\end{eqnarray}%
\begin{equation}
\Delta _{\mu }=2(\left\langle c_{j,\uparrow ,b}c_{i,\uparrow
,a}\right\rangle _{\mathrm{GS}}\delta _{\sigma \uparrow }+\left\langle
c_{j,\downarrow ,b}c_{i,\downarrow ,a}\right\rangle _{\mathrm{GS}}\delta
_{\sigma \downarrow })\delta _{\sigma \sigma ^{\prime }},
\end{equation}%
and%
\begin{equation}
E_{P}^{\mu }=(\bar{N}_{\downarrow }-\bar{N}_{\uparrow })\gamma \left\langle
P_{z}\right\rangle _{\mathrm{GS}}+\gamma \left\langle P_{z}c^{\dagger }%
\mathcal{E}_{P}^{\mu }c\right\rangle _{\mathrm{GS}}+2\gamma
\sum_{j,a}\left\langle P_{z}f_{\gamma }^{\dagger }f_{\gamma }c_{j,\downarrow
,a}^{\dagger }c_{j,\downarrow ,a}\right\rangle _{\mathrm{GS}}.
\end{equation}

The derivatives to the covariance matrix are given by Eqs. (\ref{A1})-(\ref%
{A3}) and%
\begin{eqnarray}
&&4\frac{\delta }{\delta \Gamma _{m,ij}}\left\langle P_{z}f_{\gamma
}^{\dagger }f_{\gamma }c_{l}^{\dagger }c_{k}\right\rangle _{\mathrm{GS}}
\notag \\
&=&-2\left\langle P_{z}f_{\gamma }^{\dagger }f_{\gamma }c_{l}^{\dagger
}c_{k}\right\rangle (\sqrt{1+\Theta }\frac{1}{\Gamma _{F}}\sqrt{1+\Theta }%
)_{ij}-2\left\langle P_{z}c_{l}^{\dagger }c_{k}\right\rangle _{\mathrm{GS}%
}\partial _{\theta }(\sqrt{1+\Theta }\frac{1}{\Gamma _{F}}\sqrt{1+\Theta }%
)_{ij}  \notag \\
&&+i\left\langle P_{z}\right\rangle _{\mathrm{GS}}\{\frac{1}{2}g_{F}[%
\mathcal{T}\Theta \left(
\begin{array}{c}
1 \\
-i%
\end{array}%
\right) ]_{jl}[\left( 1,i\right) \mathcal{T}^{T}]_{ki}-[\frac{1}{2}\mathcal{T%
}w(\sigma \Gamma _{m}-1)\mathcal{T}\Theta \left(
\begin{array}{c}
1 \\
-i%
\end{array}%
\right) ]_{jl}[\left( 1,i\right) \mathcal{T}^{T}]_{ki}  \notag \\
&&+[\mathcal{T}w\left(
\begin{array}{c}
1 \\
-i%
\end{array}%
\right) ]_{jl}[\left( 1,i\right) \mathcal{T}^{T}]_{ki}-\frac{1}{2}[\mathcal{T%
}\Theta \left(
\begin{array}{c}
1 \\
-i%
\end{array}%
\right) ]_{jl}[\left( 1,i\right) \mathcal{T}^{T}(\Gamma _{m}\sigma -1)w%
\mathcal{T}^{T}]_{ki}\},
\end{eqnarray}%
where%
\begin{eqnarray}
&&\partial _{\theta }(\sqrt{1+\Theta }\frac{1}{\Gamma _{F}}\sqrt{1+\Theta })
\notag \\
&=&-\frac{1}{4}w\sigma \mathcal{T}^{T}-\frac{1}{4}\mathcal{T}\sigma w+\frac{1%
}{4}\mathcal{T}\sigma w\Gamma _{m}\sqrt{1+\Theta }\frac{1}{\Gamma _{F}}\sqrt{%
1+\Theta }  \notag \\
&&+\frac{1}{4}\sqrt{1+\Theta }\frac{1}{\Gamma _{F}}\sqrt{1+\Theta }\Gamma
_{m}w\sigma \mathcal{T}^{T}+\sqrt{1+\Theta }\frac{1}{\Gamma _{F}}\sigma w%
\frac{1}{\Gamma _{F}}\sqrt{1+\Theta }.
\end{eqnarray}


\section{Ultrafast processes in RLM}

In this Appendix, we consider phototransport in the resonant level model in
the so-called infinite bandwidth limit (see ref \cite{Cuevas1990} for details).
The system Hamiltonian is given by%
\begin{equation}
H_{\mathrm{RLM}}=\sum_{k,\mathrm{a}}(k-\mu _{\mathrm{a}})c_{k,\mathrm{a}%
}^{\dagger }c_{k,\mathrm{a}}+\varepsilon _{d}d^{\dagger }d+\frac{V}{\sqrt{L}}%
\sum_{k,\mathrm{a}}(c_{k,\mathrm{a}}^{\dagger }d+\mathrm{H.c.}),
\end{equation}%
where we set $v_{\mathrm{F}}=1$, the chemical potential of the left lead is
fixed $\mu _{\mathrm{L}}=0$, and the effect of the ultrafast light is
described by $\mu _{\mathrm{R}}(t)=V_{0}\sin (\omega _{0}t)\,\theta
(t)\,\theta (T_{0}-t)$ with frequency $\omega _{0}=2\pi /T_{0}$.

In equilibrium, Green's functions of the system can be found from (see e.g.
Ref. \cite{Cuevas1990})
\begin{eqnarray}
G_{d}^{<}(\omega ) &=&i\int dte^{i\omega t}\,\langle \,0\,|d^{\dagger
}(0)d(t)|\,0\,\rangle =\frac{2i\Gamma }{(\omega -\varepsilon
_{d})^{2}+\Gamma ^{2}}n(\omega ), \\
G_{kd,\mathrm{a}}^{<}(\omega ) &=&i\int dte^{i\omega t}\,\langle
\,0\,|d^{\dagger }(0)c_{k,\mathrm{a}}(t)|\,0\,\rangle   \notag \\
&=&i\frac{V}{\sqrt{L}}n(\omega )[2\pi \delta (\omega -\varepsilon _{k,%
\mathrm{a}})\frac{1}{\omega -\varepsilon _{d}-i\Gamma }+\frac{1}{\omega
-\varepsilon _{k,\mathrm{a}}+i0^{+}}\frac{2\Gamma }{(\omega -\varepsilon
_{d})^{2}+\Gamma ^{2}}], \\
G_{kp,\mathrm{ab}}^{<}(\omega ) &=&i\int dte^{i\omega t}\,\langle \,0\,|c_{p,%
\mathrm{b}}^{\dagger }(0)c_{k,\mathrm{a}}(t)|\,0\,\rangle   \notag \\
&=&2\pi in(\omega )\delta (\omega -\varepsilon _{k\mathrm{a}})\delta
_{kp}\delta _{\mathrm{ab}}+i\frac{\Gamma }{L}n(\omega )[\frac{1}{\omega
-\varepsilon _{k\mathrm{a}}+i0^{+}}\frac{2\Gamma }{(\omega -\varepsilon
_{d})^{2}+\Gamma ^{2}}\frac{1}{\omega -\varepsilon _{p\mathrm{b}}-i0^{+}}
\notag \\
&&+2\pi \delta (\omega -\varepsilon _{p\mathrm{b}})\frac{1}{\omega
-\varepsilon _{k\mathrm{a}}+i0^{+}}\frac{1}{\omega -\varepsilon _{d}+i\Gamma
}+2\pi \delta (\omega -\varepsilon _{k\mathrm{a}})\frac{1}{\omega
-\varepsilon _{d}-i\Gamma }\frac{1}{\omega -\varepsilon _{p\mathrm{b}}-i0^{+}%
}],
\end{eqnarray}%
where $\Gamma =V^{2}$ and $n(\omega )$ is the Fermi distribution. We can use
these expressions to find expectation values of fermionic bi-linears as%
\begin{equation}
\left\langle d^{\dagger }d\right\rangle =\int_{-\infty }^{+\infty }\frac{%
d\omega }{2\pi i}G_{d}^{<}(\omega )=\int_{-\infty }^{0}\frac{d\omega }{2\pi }%
\frac{2\Gamma }{(\omega -\varepsilon _{d})^{2}+\Gamma ^{2}},
\end{equation}%
\begin{equation}
\left\langle d^{\dagger }c_{k,\mathrm{L(R)}}\right\rangle =\int_{-\infty
}^{+\infty }\frac{d\omega }{2\pi i}G_{kd,\mathrm{a}}^{<}(\omega )=\frac{V}{%
\sqrt{L}}\frac{n(k)}{k-\varepsilon _{d}-i\Gamma }+\frac{V}{\sqrt{L}}%
\int_{-\infty }^{0}\frac{d\omega }{2\pi }\frac{1}{\omega -k+i0^{+}}\frac{%
2\Gamma }{(\omega -\varepsilon _{d})^{2}+\Gamma ^{2}},
\end{equation}%
and%
\begin{eqnarray}
\left\langle c_{p,\mathrm{b}}^{\dagger }c_{k,\mathrm{a}}\right\rangle
&=&\int_{-\infty }^{+\infty }\frac{d\omega }{2\pi i}G_{kp,\mathrm{ab}%
}^{<}(\omega )  \notag \\
&=&n(k)\delta _{kp}\delta _{\mathrm{ab}}+\frac{\Gamma }{L}[\frac{1}{%
k-\varepsilon _{d}-i\Gamma }\frac{1}{k-p-i0^{+}}n(k)  \notag \\
&&+\frac{1}{p-k+i0^{+}}\frac{1}{p-\varepsilon _{d}+i\Gamma }n(p)  \notag \\
&&+\int_{-\infty }^{0}\frac{d\omega }{2\pi }\frac{1}{\omega -k+i0^{+}}\frac{%
2\Gamma }{(\omega -\varepsilon _{d})^{2}+\Gamma ^{2}}\frac{1}{\omega
-p-i0^{+}}].
\end{eqnarray}%
before the pulse arrives, where $\varepsilon _{k,\mathrm{a}}=\varepsilon _{k,%
\mathrm{b}}=k$.

When the ultrafast pulse is turned on, the current%
\begin{equation}
I=\partial _{t}\left\langle N_{\mathrm{L}}\right\rangle =\frac{2V}{\sqrt{L}}%
\sum_{k}\text{Im}\left\langle c_{k,\mathrm{L}}^{\dagger }d\right\rangle ,
\end{equation}%
i.e., the change of electron number in the left lead, can be obtained by the
Heisenberg equations of motion%
\begin{equation}
i\partial _{t}d=\varepsilon _{d}d+\frac{V}{\sqrt{L}}\sum_{k,\mathrm{a}}c_{k,%
\mathrm{a}},  \label{Ed}
\end{equation}%
and%
\begin{equation}
i\partial _{t}c_{k,\mathrm{a}}=(k-\mu _{\mathrm{a}})c_{k,\mathrm{a}}+\frac{V%
}{\sqrt{L}}d.  \label{Ec}
\end{equation}

The solution%
\begin{eqnarray}
c_{k,\mathrm{L}}(t) &=&c_{k,\mathrm{L}}(0)e^{-ikt}-i\frac{V}{\sqrt{L}}%
\int_{0}^{t}dse^{-ik(t-s)}d(s),  \notag \\
c_{k,\mathrm{R}}(t) &=&c_{k,\mathrm{R}}(0)e^{-ikt+iJ(t)}-i\frac{V}{\sqrt{L}}%
\int_{0}^{t}dse^{-ik(t-s)}e^{iJ(t)-iJ(s)}d(s),  \notag \\
d(t) &=&d(0)e^{-i(\varepsilon _{d}-i\Gamma )t}-i\frac{V}{\sqrt{L}}%
\sum_{k}c_{k,\mathrm{L}}(0)e^{-i(\varepsilon _{d}-i\Gamma
)t}\int_{0}^{t}dse^{-i(k-\varepsilon _{d}+i\Gamma )s}  \notag \\
&&-i\frac{V}{\sqrt{L}}\sum_{k}c_{k,\mathrm{R}}(0)e^{-i(\varepsilon
_{d}-i\Gamma )t}\int_{0}^{t}dse^{-i(k-\varepsilon _{d}+i\Gamma )s}e^{iJ(s)}
\end{eqnarray}%
of Eqs. (\ref{Ec}) and (\ref{Ed}) gives rise to the time dependent current%
\begin{eqnarray}
I(t) &=&-(1-e^{-2\Gamma t})\Gamma (\frac{1}{4}-\frac{1}{2\pi }\arctan \frac{%
\varepsilon _{d}}{\Gamma })  \notag \\
&&+2\Gamma ^{2}e^{-2\Gamma t}\int_{-\infty }^{0}\frac{dk}{2\pi }\text{Im}%
\frac{F_{k}(t)}{k-\varepsilon _{d}-i\Gamma }  \notag \\
&&+\Gamma ^{2}e^{-2\Gamma t}\int_{-\infty }^{0}\frac{dk}{2\pi }\left\vert
F_{k}(t)\right\vert ^{2},  \label{Ic}
\end{eqnarray}%
where the decay rate $\Gamma =V^{2}$, $J(t)=\int_{0}^{t}ds^{\prime }\mu _{%
\mathrm{R}}(s^{\prime })$, and%
\begin{equation}
F_{k}(t)=\int_{0}^{t}dse^{-i(k-\varepsilon _{d}+i\Gamma )s}e^{iJ(s)}.
\end{equation}

For the time dependent chemical potential $\mu _{\mathrm{R}}(t)$, the
function%
\begin{equation}
J(t)=\frac{V_{0}}{\omega _{0}}(1-\cos \omega _{0}t)\theta (T_{0}-t)
\end{equation}%
determines%
\begin{eqnarray}
F_{k}(t) &=&ie^{ia_{0}}\sum_{n}\frac{(-i)^{\left\vert n\right\vert
}J_{\left\vert n\right\vert }(a_{0})}{k-\varepsilon _{d}+n\omega
_{0}+i\Gamma }\{[e^{-i(k-\varepsilon _{d}+n\omega _{0}+i\Gamma )t}-1]\theta
(T_{0}-t)  \notag \\
&&+[e^{-i(k-\varepsilon _{d}+n\omega _{0}+i\Gamma )T_{0}}-1]\theta
(t-T_{0})\}  \notag \\
&&+i\frac{e^{-i(k-\varepsilon _{d}+i\Gamma )t}-e^{-i(k-\varepsilon
_{d}+i\Gamma )T_{0}}}{k-\varepsilon _{d}+i\Gamma }\theta (t-T_{0})
\end{eqnarray}%
by the expansion%
\begin{equation}
e^{-ia_{0}\cos \omega _{0}s}=\sum_{n}(-i)^{\left\vert n\right\vert
}J_{\left\vert n\right\vert }(a_{0})e^{-in\omega _{0}s},
\end{equation}%
where $a_{0}=V_{0}/\omega _{0}>0$.

The number of electrons transport from the left lead to the right one is
defined as $N_{\mathrm{tran}}=\int_{0}^{\infty }dtI(t)$. The integral of $%
I(s)$ over time gives%
\begin{eqnarray}
N_{\mathrm{tran}} &=&(\frac{1}{2}-\Gamma T_{0})(\frac{1}{4}-\frac{1}{2\pi }%
\arctan \frac{\varepsilon _{d}}{\Gamma })  \notag \\
&&+2\Gamma ^{2}\int_{-\infty }^{0}\frac{dk}{2\pi }\text{Im}\frac{C_{1}(k)}{
k-\varepsilon _{d}-i\Gamma }+\Gamma ^{2}\int_{-\infty }^{0}\frac{dk}{2\pi }%
C_{2}(k),  \label{Nt}
\end{eqnarray}%
where%
\begin{eqnarray}
C_{1}(k) &=&ie^{ia_{0}}\sum_{n}\frac{(-i)^{\left\vert n\right\vert
}J_{\left\vert n\right\vert }(a_{0})}{k-\varepsilon _{d}+n\omega
_{0}+i\Gamma }[i\frac{e^{-i(k-\varepsilon _{d}+n\omega _{0}-i\Gamma )T_{0}}-1%
}{k-\varepsilon _{d}+n\omega _{0}-i\Gamma }  \notag \\
&&+\frac{1}{2\Gamma }e^{-i(k-\varepsilon _{d}+n\omega _{0}-i\Gamma )T_{0}}-%
\frac{1}{2\Gamma }]  \notag \\
&&+\frac{e^{-i(k-\varepsilon _{d}-i\Gamma )T_{0}}}{(k-\varepsilon
_{d})^{2}+\Gamma ^{2}}-i\frac{1}{2\Gamma }\frac{e^{-i(k-\varepsilon
_{d}-i\Gamma )T_{0}}}{k-\varepsilon _{d}+i\Gamma },
\end{eqnarray}%
and%
\begin{eqnarray}
C_{2}(k) &=&\sum_{nm}\frac{(-1)^{\left\vert n\right\vert }i^{\left\vert
n\right\vert +\left\vert m\right\vert }J_{\left\vert n\right\vert
}(a_{0})J_{\left\vert m\right\vert }(a_{0})}{(k-\varepsilon _{d}+n\omega
_{0}+i\Gamma )(k-\varepsilon _{d}+m\omega _{0}-i\Gamma )}  \notag \\
&&[i\frac{e^{-i(n-m)\omega _{0}T_{0}}-1}{(n-m)\omega _{0}}-i\frac{
e^{-i(k-\varepsilon _{d}+n\omega _{0}-i\Gamma )T_{0}}-1}{k-\varepsilon
_{d}+n\omega _{0}-i\Gamma }  \notag \\
&&-\frac{e^{i(k-\varepsilon _{d}+m\omega _{0}+i\Gamma )T_{0}}-1}{
i(k-\varepsilon _{d}+m\omega _{0}+i\Gamma )}-\frac{e^{-2\Gamma T_{0}}-1}{
2\Gamma }]  \notag \\
&&+\frac{e^{-2\Gamma T_{0}}}{2\Gamma }\left\vert C_{3}(k)\right\vert ^{2}+2%
\text{Re}\frac{ie^{i(k-\varepsilon _{d}+i\Gamma )T_{0}}}{(k-\varepsilon
_{d})^{2}+\Gamma ^{2}}C_{3}(k)
\end{eqnarray}%
is determined by%
\begin{equation}
C_{3}(k)=e^{ia_{0}}\sum_{n}\frac{(-i)^{\left\vert n\right\vert
}J_{\left\vert n\right\vert }(a_{0})}{k-\varepsilon _{d}+n\omega
_{0}+i\Gamma }[e^{-i(k-\varepsilon _{d}+n\omega _{0}+i\Gamma )T_{0}}-1]-%
\frac{e^{-i(k-\varepsilon _{d}+i\Gamma )T_{0}}}{k-\varepsilon _{d}+i\Gamma }.
\end{equation}

For RTM with the tight-binding dispersion relation, it is difficult to
obtain the transient current and the transport number $N_{\mathrm{tran}}$
analytically. One can investigate the dynamics governed by the quadratic
Hamiltonian $H_{\mathrm{RLM}}$ numerically. Here, we consider the pulse
profile (\ref{Ve}) in the main text. The transient current is shown in Figs. %
\ref{RTMtb}a and \ref{RTMtb}b for $\varepsilon _{d}=-0.5$ and $\varepsilon
_{d}=0$, where $\Gamma =0.25$, $\omega _{0}=1$, and $V_{0}=0.5$, $1$, $1.5$,
and $2$. The transport electron number $N_{\mathrm{tran}}$ as a function of $%
V_{0}$ is shown in Figs. \ref{RTMtb}c and \ref{RTMtb}d.

\begin{figure}[tbp]
\includegraphics[width=0.9\linewidth]{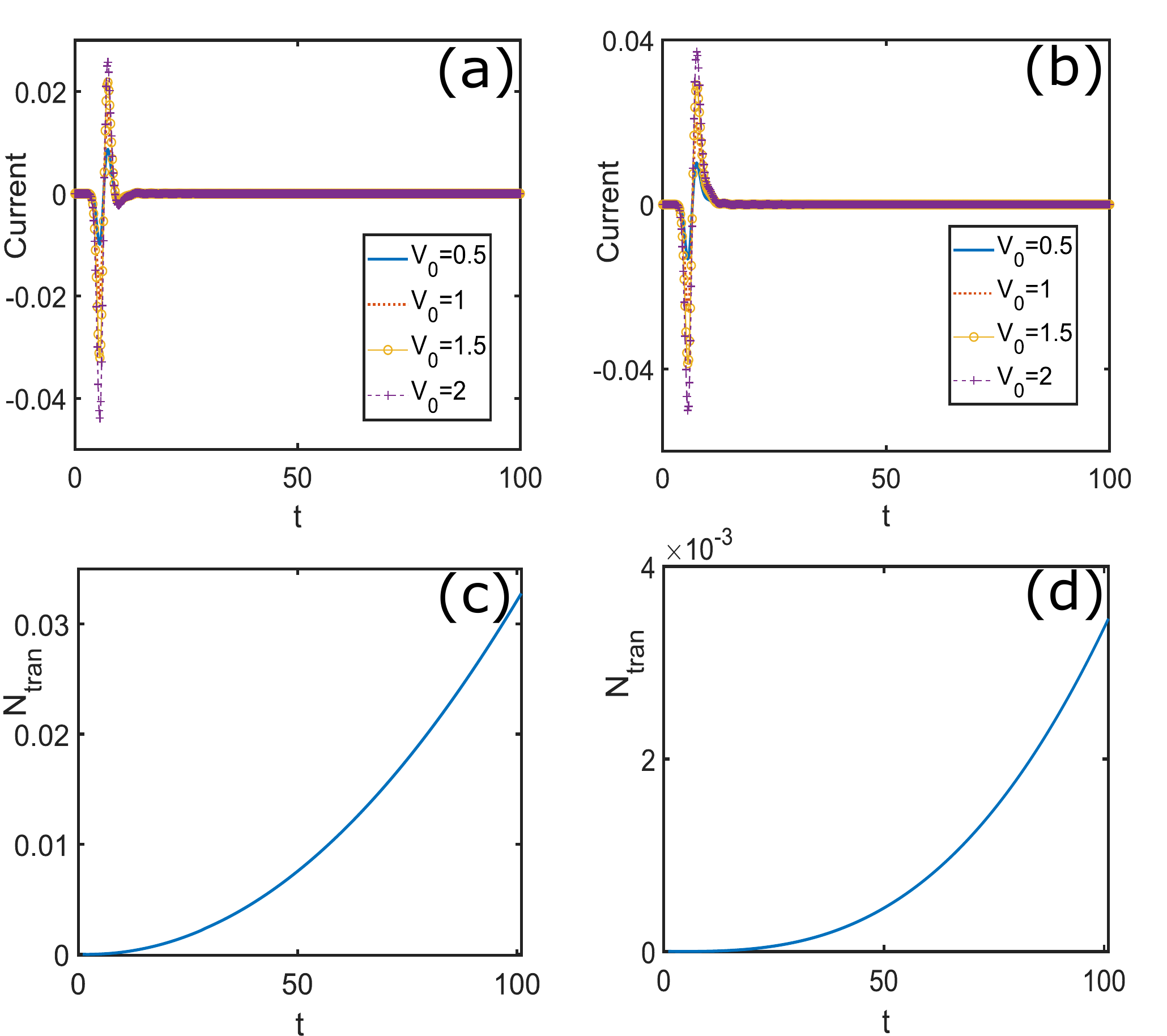}
\caption{Ultrafast dynamics of RTM with the tight-binding dispersion
relation, where $\Gamma =0.25$ and $\protect\omega _{0}=1$. (a)-(b) The
transient current for $\protect\varepsilon _{d}=-0.5$ and $\protect%
\varepsilon_{d}=0$; (c)-(d) The transport electron number $N_{\mathrm{tran}}$
for $\protect\varepsilon _{d}=-0.5$ and $\protect\varepsilon_{d}=0$.}
\label{RTMtb}
\end{figure}

\end{widetext}

\end{document}